%% file: main_arxiv.tex
\documentclass[acmtog,authorversion,printacmref=false]{acmart}
\setcitestyle{square}

\setcopyright{none}
\acmYear{}
\copyrightyear{2018}
\acmArticle{}
\acmMonth{}
\acmVolume{}
\acmNumber{}
\acmDOI{}
\input{my_macros.inc}

\begin{document}
\title[\name: Interaction-guided Joint Scene and Human Motion Mapping from Monocular Videos]{\name: Interaction-guided Joint Scene and Human Motion Mapping from Monocular Videos}  

\author{Aron Monszpart, Paul Guerrero, Duygu Ceylan, Ersin Yumer, Niloy J. Mitra}
\renewcommand{\shortauthors}{Aron Monszpart, Paul Guerrero, Duygu Ceylan, Ersin Yumer, Niloy J. Mitra}

\input{00_abstract}

\if0
\fi


\begin{teaserfigure}
  \begin{overpic}[width=\textwidth]{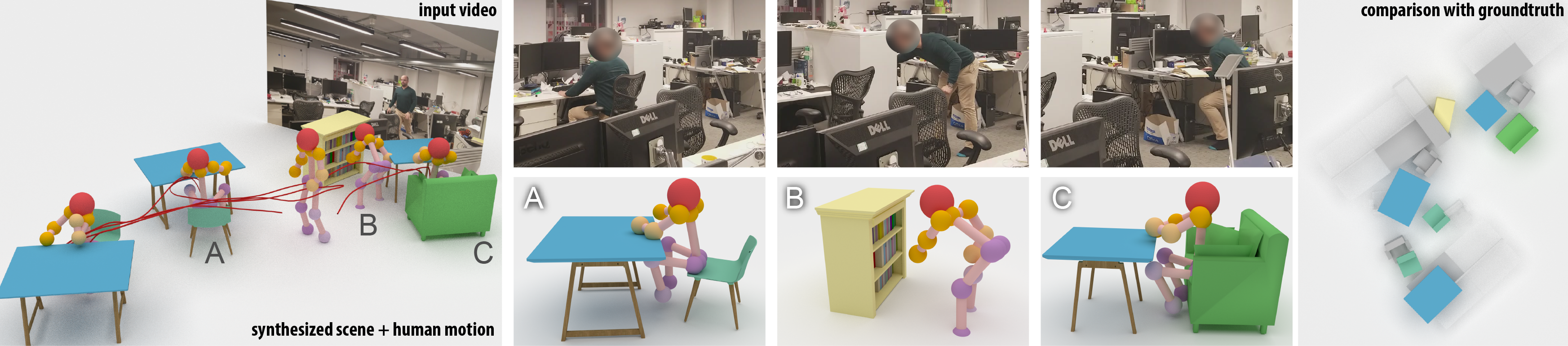}
  	\Large
  	\put(92.5,.5){\color{black}\textbf{\Scene{office1-1}}}
  \end{overpic}
  \caption{We present \name, a method that reasons about the interactions of humans with objects, to recover both a plausible scene arrangement and human motions, that best explain an input monocular video (see inset). We fit characteristic interactions called \emph{scenelets} (\eg A, B, C) to the video and use them to reconstruct a plausible object arrangement and human motion path (left). The key challenge is that reliable fitting requires information about occlusions, which are unknown (\ie latent). (Right)~We show an overlay (from top-view) of our result over manually annotated groundtruth object placements. Note that object meshes are placed based on estimated object category, location, and size information. }
  \label{fig:teaser}
\end{teaserfigure}

\maketitle

\input{01_introduction}

\input{02_relatedWork}

\input{03_overview}

\input{04_method}

\input{060_evaluation}

\input{07_discussion}

\bibliographystyle{ACM-Reference-Format}
\bibliography{geo_references,sceneActionCoupled}

\appendix
\input{appendix}

\end{document}

%% file: my_macros.inc
\usepackage{booktabs} % For formal tables
\usepackage{enumerate}

\usepackage{etoolbox}
\makeatletter
\patchcmd\@combinedblfloats{\box\@outputbox}{\unvbox\@outputbox}{}{%
  \errmessage{\noexpand\@combinedblfloats could not be patched}%
}%
\makeatother
\usepackage[normalem]{ulem} % \sout for strikethrough
\usepackage{amsmath}

\usepackage{xspace}
\newcommand{\name}{i\textsc{Mapper}\xspace}
                                                                                                         
\newlength{\myparaskip}
\newcommand{\mypara}[1]{\vspace{\myparaskip} \noindent \textbf{#1}} % originally: \vspace{0.05in}
\usepackage[percent]{overpic}
\newcommand{\vnudge}{\vspace{-.1in}}
\def\ie{{\it i.e.,\ }}

\def\eg{{\it e.g.,\ }}

\def\cf{{\it cf.,\ }}

\newcommand{\revision}[1]{#1}

\DeclareMathOperator*{\argmin}{\arg\!\min}

\newcommand{\best}[1]{\textbf{#1}}
\newcommand{\lcrnet}{\mbox{LCR-Net++}\xspace}
\newcommand{\lcrnetthreed}{\mbox{LCR-Net++3D}\xspace}
\newcommand{\tomethreed}{\mbox{Tome3D}\xspace}
\newcommand{\frcnn}{\mbox{FRCNN}\xspace}
\newcommand{\maskrcnn}{\mbox{Mask-RCNN}\xspace}
\newcommand{\xiactort}[3]{\xi_{#1 \rightarrow #2}^{#3}}
\newcommand{\xiactor}{\xiactort{a}{s}{t}}
\newcommand\bigforall{\mbox{\Large $\mathsurround0pt\forall$}} 

\graphicspath{{./images/}}
\definecolor{turquoise}{cmyk}{0.65,0,0.1,0.1}
\definecolor{purple}{rgb}{0.65,0,0.65}
\definecolor{dark_green}{rgb}{0, 0.5, 0}
\definecolor{orange}{rgb}{0.8, 0.2, 0.2}
\definecolor{red}{rgb}{1, 0, 0}

\newcommand{\cma}[1]{\textcolor[rgb]{0.35,0.56,0.41}{\textbf{\small A: #1}}}
  
\newcommand{\ersin}[1]{\textcolor[rgb]{0.1,0.5,0.2}{\textbf{Ersin: #1}}}

\let\hl\highlightSA

\newcommand{\DefineScene}[2]{%
	\expandafter\newcommand\csname rmk-#1\endcsname{#2}%
}
\newcommand{\Scene}[1]{\csname rmk-#1\endcsname}

\DefineScene{office1-1}{Office1}
\DefineScene{Office1}{\Scene{office1-1}}
\DefineScene{office2-1}{Office2}
\DefineScene{Office2}{\Scene{office2-1}}
\DefineScene{livingroom00}{LvRoom}
\DefineScene{Livingroom}{LvRoom}
\DefineScene{lobby15}{Lobby1}
\DefineScene{Lobby1}{Lobby1}
\DefineScene{lobby22-1}{Lobby2}
\DefineScene{Lobby2}{Lobby2}
\DefineScene{lobby11}{Scene1}
\DefineScene{Lobby3}{\Scene{lobby11}}
\DefineScene{Scene1}{\Scene{lobby11}}
\DefineScene{fig8row1}{\Scene{lobby11}}
\DefineScene{lobby12}{Scene2}
\DefineScene{Lobby4}{\Scene{lobby12}}
\DefineScene{Scene2}{\Scene{lobby12}}
\DefineScene{fig8row2}{\Scene{lobby12}}
\DefineScene{lobby18-1}{Scene3}
\DefineScene{Lobby5}{\Scene{lobby18-1}}
\DefineScene{Scene3}{\Scene{lobby18-1}}
\DefineScene{fig8row3}{\Scene{lobby18-1}}
\DefineScene{lobby19-3}{Scene4}
\DefineScene{Lobby6}{\Scene{lobby19-3}}
\DefineScene{Scene4}{\Scene{lobby19-3}}
\DefineScene{fig8row4}{\Scene{lobby19-3}}
\DefineScene{lobby24-3-3}{Scene5a} % one sit
\DefineScene{Lobby7a}{\Scene{lobby24-3-3}}
\DefineScene{Scene5a}{\Scene{lobby24-3-3}}
\DefineScene{fig10tl}{\Scene{lobby24-3-3}}
\DefineScene{lobby24-3-1}{Scene5b} % two sits
\DefineScene{Lobby7b}{\Scene{lobby24-3-1}}
\DefineScene{Scene5b}{\Scene{lobby24-3-1}}
\DefineScene{fig10tr}{\Scene{lobby24-3-1}}
\DefineScene{lobby24-3-2}{Scene5c} % three sits
\DefineScene{Lobby7c}{\Scene{lobby24-3-2}}
\DefineScene{Scene5c}{\Scene{lobby24-3-2}}
\DefineScene{fig10bl}{\Scene{lobby24-3-2}}
\DefineScene{lobby24-2-2}{Scene5d} % reverse three sits
\DefineScene{Lobby7d}{\Scene{lobby24-2-2}}
\DefineScene{Scene5d}{\Scene{lobby24-2-2}}
\DefineScene{fig10br}{\Scene{lobby24-2-2}}
\DefineScene{garden1}{Garden}
\DefineScene{Garden}{Garden}
\DefineScene{library3}{Library1}
\DefineScene{Library1}{\Scene{library3}}

\DefineScene{library1}{Library2}
\DefineScene{Library2}{\Scene{library1}}

\DefineScene{angrymen00}{Angrymen}
\DefineScene{Angrymen}{\Scene{angrymen00}}

\DefineScene{grease00-11}{Grease}
\DefineScene{Grease}{\Scene{grease00-11}}

\DefineScene{vnect1}{VNect1} % cam2
\DefineScene{vnect2}{VNect2} % cam2-2
\DefineScene{vnect}{GrRoom} % greenRoom

\usepackage{cleveref} % has to come after hyperref
\Crefname{figure}{Fig.}{Figs.}

%% file: 00_abstract.tex
\begin{abstract}

A long-standing challenge in scene analysis is the recovery of scene arrangements under moderate to heavy occlusion, directly from monocular video. 
While the problem remains a subject of active research, concurrent advances have been made in the context of human pose reconstruction from monocular video, including image-space feature point detection and 3D pose recovery. %Notable advances include image-space feature point detection, local 3D pose recovery, and global pose recovery. 
These methods, however, start to fail under moderate to heavy occlusion as the problem becomes severely under-constrained. 
We approach the problems differently. We observe that people \textit{interact similarly in similar scenes}. Hence, we exploit the correlation between scene object arrangement and motions performed in that scene in both directions: first, typical motions performed when interacting with objects inform us about possible object arrangements; and second, object arrangements, in turn, constrain the possible motions.
%under similar scene arrangements, humans interact with objects {\em similarly} (e.g., sitting down on a sofa, picking up a book from a shelf, etc.).
%

We present \name, a data-driven method  that focuses on identifying human-object interactions, and {\em jointly} reasons about objects and human movement over space-time to recover both a plausible scene arrangement and consistent human interactions. 
We first introduce the notion of {\em characteristic interactions} as regions in space-time when an informative human-object interaction happens. This is followed by a novel {\em occlusion-aware matching} procedure that searches and aligns such characteristic snapshots from an interaction database to best explain the input monocular video. 
Through extensive evaluations, both quantitative and qualitative, we demonstrate that \name significantly improves performance over both dedicated state-of-the-art scene analysis and 3D human pose recovery approaches, especially under medium to heavy occlusion.

\end{abstract}

\if0

3D geometry estimation from monocular video is a severely ill-posed problem due to the lack of relialably identifiable depth cues. 
Similarly, global 3D pose estimation of human motion is under-constrained due to depth ambiguity and foreshortening.
In this paper we present a method, where reasoning about both geometry and human motion in a coupled fashion allows for creating better estimates compared to other methods in the respective fields.
The key idea is, that human poses change smoothly over time when performing easy-to-identify actions, and convey strong semantic cues about the surrounding environment, which in turn helps disambiguating the global embedding of the poses.
We evaluate compared to state-of-the-art methods in both fields, and present interesting use cases.
\fi

%% file: 01_introduction.tex
\section{Introduction} \label{sec:introduction}

\if0
\begin{figure}[h]
	\begin{overpic}[width=\columnwidth]{placeholders/Office2.png}
		\Large
		\put(2, 47){\color{white}\textbf{\Scene{office2-1}}}
	\end{overpic}
	\caption{possible alternate teaser.}
	\label{f:office2}
\end{figure}
\fi

%para1
%Separate options for capturing detailed geometric scenes and motion/performance. On the one hand, various options for capturing complex scenes -- kinect fusion, bundle fusion, etc. but requires camera to be moved around to 'see around' occlusion. On the other hand, various options for capturing motion/performance but heavily relies on human templates that are matched without occlusions. Typically they are treated separately.  We are interested in jointly capturing both plausible scene layouts and performance from monocular videos.
%
Digitizing the physical world is critical for  many emerging fields such as virtual and augmented reality, smart home systems, or robotics. 
Such applications require access to not only reconstructions of the physical spaces, but also an {\em understanding} of the common human actions performed in such spaces. For example, our future personal robot assistants should not only know the object layout of our lounges, but also our working habits in these spaces. 
%

\begin{figure}[t!]
  \includegraphics[width=\columnwidth]{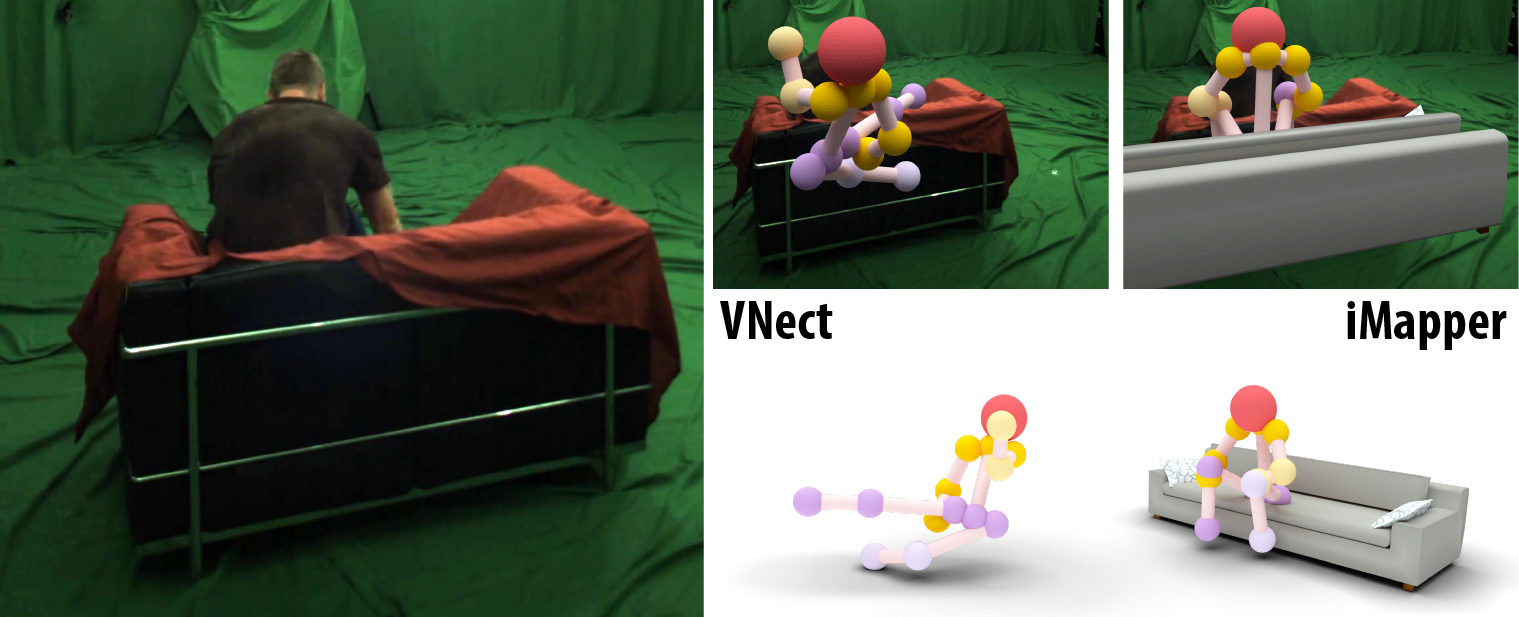}
  \caption{Comparison of state-of-the-art 3D human pose detection from monocular video VNect~\protect\cite{VNect_SIGGRAPH2017} with \name. Note how VNect breaks down in regions of occlusion (VNect was not designed to handle occlusions), while our approach continues to produce plausible results because of explicit occlusion detection and handling. Bottom row shows the recovered human pose from another camera view for better visibility. 
  }
  \label{fig:vnect comparison}
\end{figure}

%para2
Traditionally, researchers have tackled scene estimation and human performance capture as two separate problems. On the one hand, scene reconstruction methods such as Kinect Fusion~\cite{Newcombe:2011:KRD} and Bundle Fusion~\cite{dai2017bundlefusion} can produce high-quality static indoor reconstructions, and the likes of DynamicFusion~\cite{Newcombe2015DynamicFusionRA} can capture non-rigidly deforming scenes. These methods require the sensor to be manually moved to see around occlusions making the capture process cumbersome. Moreover, the process needs to be repeated each time scene objects are moved. 
On the other hand, there are various options for reconstructing 3D human performances, either using multiple sensors~\cite{vonMarcard:2017} or monocular video input, based on a CNN pose regressor along with kinematic skeleton fitting~\cite{VNect_SIGGRAPH2017}. These methods assume performances to be free from object-induced occlusions (see Figure~\ref{fig:vnect comparison} and Section~\ref{sec:evaluation}). 

%a CNN-based pose regressor along with kinematic skeleton fitting on monocular video input~\cite{VNect_SIGGRAPH2017}, but assume performances to be free from object-induced occlusions (see Figure~\ref{fig:vnect comparison}). %Other deep learning based approaches consume only images or videos as input, but utilize training data that have been mostly captured in controlled environments~\cite{h36m_pami,andriluka14cvpr}. 

%para 3
While indoor scene configurations can be extremely rich and diverse, we observe that a large fraction of them are linked by a common thread --- {\em they are regularly inhabited  by humans}. Moreover, in similar scene configurations, social behavioral rules lead to humans typically performing similar actions (\cf \cite{principles13}). Examples of such actions include sitting on sofas, picking up books from shelves, or walking around obstacles. Hence, instead of tackling scene modeling and human reconstruction as separate problems, we propose to {\em jointly} estimate both plausible scene layouts and consistent human performances from monocular video.

%para 4
%The main challenge are occlusions / not directly observed objects and performance: 1) how to detect objects that are strongly / fully occluded from the camera? 2) How to detect strongly occluded performance? 1) has been targetted by joint semantic segmentation + completion (from RGBD data?) or using physical stability arguments (imagining the unseen). 2) has not been done as far as we know?
%
A fundamental challenge in 3D reconstruction that can benefit from such an approach is working with scenes under {\em occlusion}. 
% since this is not only a challenge with our approach, but with all 3D reconstruction, and our approach is especially suited to tackle it
%A fundamental challenge in reaching the above goal is working with scenes under {\em occlusion}. 
%A fundamental challenge in acquiring 3D representations of the physical world is due to occlusions. 
A successful solution needs to tackle two  problems: (i)~How to hallucinate information about partially or fully hidden scene objects in monocular input? (ii)~How to determine human performance that is strongly occluded by various (unknown) scene objects? 
% simpler?: (ii) How to identify a human performance that interacts with occluded scene objects?
%  
It is believed that, as humans, we focus on the interactions of actors with the objects in the scene (referred to as {\em anticipation} in \cite{psyc76}), instead of separately identifying objects and human performances.
Our experience allows us to compensate for missing information in {\em both} objects and performances, using subtle hints arising from their interaction.
%Over the years, we develop an understanding of the scene context and configuration as well as human performance. In other words, by collecting subtle hints arising due to human-object interactions, we compensate for missing information in both the channels.
For example, in the video for the scene shown in Figure 1, we can `see' the person walk behind the desk and sit down -- from that, we can imagine both the person's sitting pose over time \emph{and} the location of the unseen chair; similarly, for the person picking up an object from the shelf (see also supplementary video).  
%along with its pose.
%In the video for the scene shown in Figure 1(b), we can infer the actor's `sitting down' performance from just seeing the actor's head and the couch. Exchanging the couch with a bed, we might have inferred a `lying down' performance instead.
%
%The idea of interaction has been explored in PiGraphs in the forward setting. We focus on the reverse direction ... 

%para5
We propose \name, a data driven method that
%masters a similar perception
accomplishes a similar feat
by jointly reasoning about interaction detection and object placement. We rely on humans behaving consistently near similar object configurations at different times. Hence, as a data-prior, we use a database of 3D interactions (extracted from the PiGraph dataset~\cite{savva2016pigraphs}) between humans over time and local object configurations that we call \emph{scenelets}.
%\cma{Aron: I put in the citation for PiGraphs here instead of xx, I assume that's what you meant.}
Human actions are easier to robustly detect than objects using state-of-the-art methods, and the scenelets in our dataset give us typical object configurations associated with actions. Additionally, a scenelet also tells us which parts of human actors are likely to be occluded when viewed from different angles.
%, which we can compare with observed occlusions.

%para6
Given a monocular video, we fit such scenelets to each actor's visible 2D joints, and ensure that the joints missing in the video match the occlusions we expect from the candidate scene arrangement. 
However, we have a cyclic dependency: 
% the matching of the recovered scene objects relies
the recovered scene objects depend
on the quality of the human pose estimates; while, the estimation of human poses, in turn, needs to know which 2D feature points remain unoccluded by the discovered scene objects.  
Hence, we first generate candidate scenelets matching potentially informative segments of the video, and then solve a global selection problem to extract a consistent set of scenelets fitted to the observed interactions. This simultaneously produces  an unoccluded 3D human performance and a set of static objects that plausibly fit the observed interactions (see Figure~\ref{fig:teaser}). 

%para6
We extensively evaluated \name\ on a range of scenes of varying complexity and occlusion. We provide both qualitative and quantitative evaluation on real data demonstrating that our simultaneous estimation (\ie  scene layout and human pose) improves upon dedicated state-of-the-art methods that  treat the two problems independently. 
In summary, our main contributions are: 
(i)~proposing the first method that jointly reasons about 3D scene modeling and plausible human object interactions given monocular video input of scenes with occlusion; 
(ii)~detecting which human motions in a sequence are
%parts of a human interaction is
informative, when such \emph{characteristic} pose sequences occur, and matching them to the scenelet database while accounting for (unknown) occlusion; and
(iii)~combining detected scenelets into a consistent 3D scene and human performance by a novel optimization that reasons about several cues including number of objects and object categories, position and orientation of objects, realistic room layouts, and compatible human movements.

%% file: 02_relatedWork.tex
\begin{figure*}[t!]
    \centering
    \includegraphics[width=\textwidth]{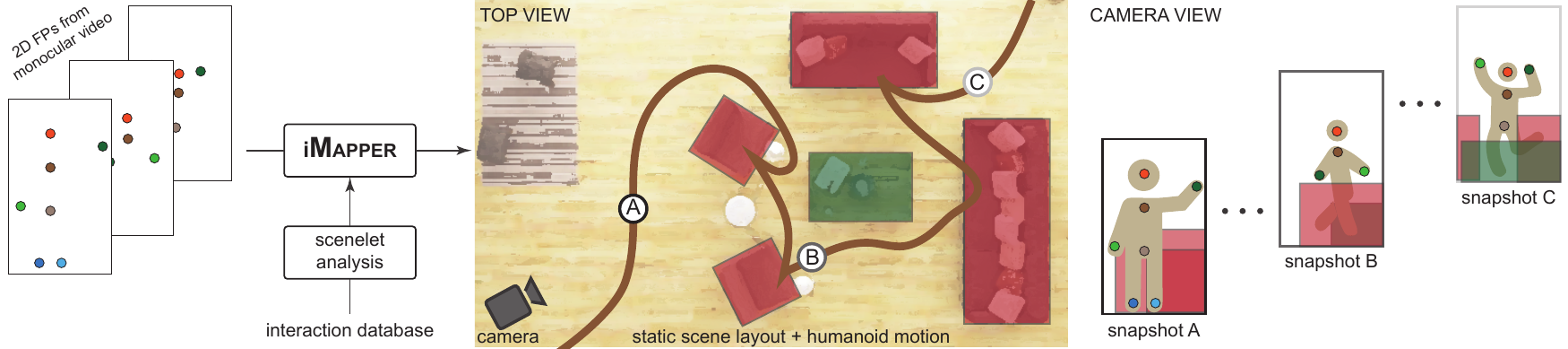}
    \caption{Starting from a monocular video and an interaction database, \name\ produces a {\em plausible} and {\em consistent} static scene layout and humanoid 3D motion path (shown as brown curve) over time. In a plausible solution, the projection of the discovered human motion under the input camera agrees with per frame 2D feature points in the unoccluded parts. In a consistent solution, we expect the scene and human motions to have matching interactions in space-time. By jointly reasoning over occlusions and interactions, we are able to improve both scene layout (\eg depth estimation) and pose estimation (\eg robustness under latent occlusion). For example, in this illustration, the inferred woman's pose  remains unoccluded in `snapshot A', while she is partly occluded in `snapshot B' and `snapshot C' by inferred scene objects.   }
    \label{fig:imapper_illustration}
\end{figure*}

\section{Related work} \label{sec:related_work}
Our system is related to prior work on scene analysis and synthesis, human-centric shape analysis, as well as human pose estimation. We now discuss selected papers in these categories to better position our approach.

 %\niloy{integrate this .. } To tackle the first problem, researchers have used regularization priors such as retrieving database objects to complete RGBD scans~\cite{kmyg_acquireIndoor_sigga12}, intra-object similarity~\cite{xx}, inter-object relations~\cite{xx}, and scene statistics~\cite{xx}. There is relatively less work that tackles the second question. Existing solutions either approach this problem in the context of 2D pose~\cite{Fu_2015_ICCV}, or represent 3D human pose as sparse linear  combination of known 3D poses and recover these blend weights from partially observed data~\cite{Huang:2009}. Such approaches, however, are limited to image input, do not consider the temporal dynamics of the human motion, and do not reason explicitly about occlusion arising due to human-object interactions.

\mypara{Scene analysis and synthesis.} With the advances in acquisition technologies, several large scale indoor reconstruction datasets have been created~\cite{Matterport3D,dai2017scannet}. Such datasets, together with the availability of 3D scene collections, are attracting more attention to 3D scene analysis. Several previous works focus on analyzing inter-object relationships~\cite{fisher2011,Zhao:2016,HUANG201646} and hierarchical grammars~\cite{Liu:2014} given a 3D scene collection. Xu et al.~\shortcite{Xu:2014} introduce the concept of \emph{focal points} that represent frequently occurring sub-arrangements of objects across heterogeneous collections of scenes. For synthesis purposes, one line of related work focuses on modeling scenes from single images~\cite{Satkin2013,Poirson2016,izadinia2017im2cad} or RGBD scans~\cite{Shao:2012,Nan:2012} by matching individual 3D objects. Other work~\cite{Schwing2013,Chen:2014} further explores spatial relationships between objects. Given a room layout and object instances, Yeh et al.~\shortcite{Yeh:2012} provide a Markov Chain Monte Carlo based algorithm to synthesize object arrangements that take spatial constraints into consideration. Fisher et al.~\shortcite{2012-scenesynth} present an example-based synthesis approach that uses a probabilistic graphical model to encode object relationships. Del Pero et al.~\shortcite{Pero_2013_CVPR} represent 3D objects as a collection of primitives to reconstruct more accurate room geometry from images.

\mypara{Human-centric scene synthesis.} Recent work in scene synthesis incorporates human actions into scene analysis (i.e., object arrangements, scene layout) to infer where specific actions can take place~\cite{savva2014scenegrok} or where a new object may be placed~\cite{Jiang:2016} in a scene. Frank et al.~\shortcite{Frank2015} recognize certain human actions to insert 3D objects into a SLAM-based reconstruction output. Ma et al.~\shortcite{Ma:2016:AIS:2980179.2980223} model both object-object and object-human actions to refine an input 3D scene. The recent work of Fu et al.~\shortcite{fu_siga17} analyzes a collection of floor plans annotated with possible human actions and 3D objects to guide the synthesis of a scene given an empty room layout and a few object categories. In contrast to these approaches, our work does not assume any initial input about the scene geometry or layout.

One of the earlier works that uses motion cues to reason about the scene geometry is the work of Brostow and Essa~\shortcite{brostow1999motion}. Given an input video, this work represents motion in a general sense by segmenting each video frame into blobs that are classified as static or active. This classification is then used to extract depth layers for the scene yielding a 2.5D representation. %In contrast, our goal is to synthesize a 3D scene by identifying the object types and their arrangement that can explain the observed motion. 
Fouhey et al.~\shortcite{Fouhey:PeopleWatching:12} combine human pose estimates with appearance and other geometric cues to estimate the room layout and free space in a single image or a time-lapse video. Compared to these approaches, we focus on a fundamentally different problem: instead of recovering a faithful geometry reconstruction, our goal is to generate a plausible scene as well as a human motion that that can explain the input video.% Thus, instead of relying on appearance information we combine motion cues with scene statistics learned from a database of 3D scenes.

One of the previous systems most closely related to our approach is the work of Fisher et al.~\shortcite{fisher2015actsynth} that utilizes a scene template computed from a given 3D scan, together with an activity classifier, to model plausible layouts. There are two main differences between our approach and this work. First, they require an initial 3D scan of the scene to be provided in order to predict possible actions that can take place in the scene. In contrast, the input to our system is the human motion sequence with no prior knowledge of the scene geometry. Second, instead of operating at the level of high-level activity labels, which rely on a pre-defined set of activity classes, we use \emph{scenelets}, \ie short human pose-object interaction sequences, where any pose example provides a cue for the scene geometry, \ie presence of a specific object or void space. While we bootstrap the scene population process by identifying \emph{characteristic poses} that imply the presence of typical objects, any input pose provides constraints for identifying the occupied and void regions of the scene.

The recent work of Savva et al.~\shortcite{savva2016pigraphs} analyzes interaction snapshots, \ie action and pose labeled RGBD sequences to learn \emph{prototypical interaction graphs (PiGraphs)} which link the attributes of the human pose to the surrounding object geometries. They show how PiGraphs can be utilized to generate scenes that correspond to static interaction snapshots (\eg lie on bed). In contrast, we focus on observing a dynamic motion sequence that consists of different actions. We combine \emph{scenelets} which depict short sequences of human actions to synthesize a scene that is in agreement for the whole duration of the motion. Finally, Kang et al.~\shortcite{KANG201725} focus on a similar goal of scene synthesis that explores motion cues. However, their input is a 3D human motion sequence free of occlusions. In contrast, we use monocular videos with moderate to severe occlusions as input. Thus, our goal is not only to synthesize plausible scene objects but also recover the 3D human motion where occlusions happen.

\begin{figure*}[t!]   
\includegraphics[width=\textwidth]{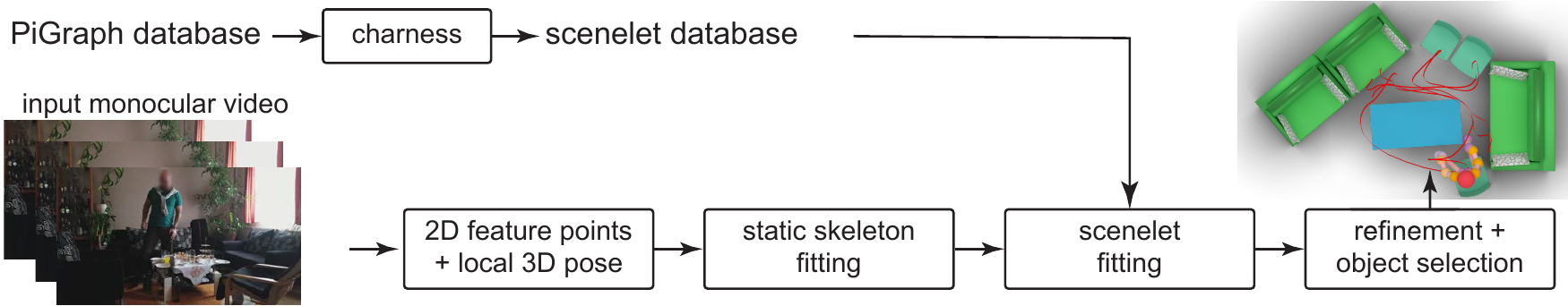}
  \caption{Overview of our approach. Please refer to the text for details of the individual steps.  }
  \label{fig:overview}
  \vnudge
\end{figure*}

\mypara{Human-centric shape analysis.} Earlier work that uses observations of how humans interact with objects focuses on tasks such as object and event recognition~\cite{Gupta:2009:OHI:1608576.1608766,delaitre2012,Wei:2013} and action detection~\cite{icml11_yao}. Kim et al.~\shortcite{Kim:2014:SHS} propose a shape analysis tool based on a human-object affordance model that can be used for many applications including correspondence estimation, shape retrieval, and view selection. In a  followup project, Fu et al.~\shortcite{Fu:2017tvcg} utilize a similar model to generate new objects by combining functional parts of existing objects. The recent work of Pirk et al.~\shortcite{Pirk:2017}, introduces the concept of \emph{interaction landscapes} which provides a descriptor of an object based on the type of interactions it can be involved in. In a follow-up paper~\cite{Pirk:2017icip}, they extend this notion to compute descriptors for individual interactions. More recently, Gkioxari et al.~\shortcite{gkioxari2017} predict human-verb-object instances from a single image to characterize human-object interactions. In our work, on the other hand, instead of focusing on individual object-based inference tasks, we use observed human motion to generate plausible scenes from a diverse and rich set of possibilities.

\mypara{Human pose estimation.} %There is a large body of work on both 2D and 3D human pose estimation from images and videos. In the context of 2D pose estimation, earlier work uses part detectors and graphical models~\cite{Andriluka:2009}. 
With the recent success of deep learning, we have seen advances both in 2D~\cite{Toshev:2014,Newell:PoseHg:2016,Insafutdinov:DeeperCut:2016,wei2016,cao2017} and 3D pose estimation~\cite{HuangKeyframes2014CVPR,Zhou:Monocap:2015,Tekin2016,tome2017lifting,RogezWS18}. In particular, the recent VNect system~\cite{VNect_SIGGRAPH2017} demonstrates state-of-the-art results for global pose estimation from monocular video.
%
%both in 2D convolutional neural networks (CNNs) have been utilized for body part detection and association~\cite{Tompson:2014,Insafutdinov:DeeperCut:2016,cao2017}. Other methods use deep networks to infer 2D joint locations~\cite{Toshev:2014,Newell:PoseHg:2016,wei2016}. A common strategy for 3D pose estimation is to lift 2D joint predictions to 3D. While some methods represent 3D pose as a linear combination of sparse pose basis and estimate basis coefficients~\cite{ramakrishna2012,Zhou:Monocap:2015} other methods retrieve the closest 3D pose from a set of examples given the 2D pose estimation~\cite{Yasin2016}. Du et al.~\shortcite{Du:MarkerLess:2016} lift each 2D joint to 3D by incorporating temporal smoothness constraints on the movement speed of the joints. In case of video input, several works have explored spatio-temporal features~\cite{Tekin2016} and optical flow constraints~\cite{Alldieck:2017} to infer 3D pose directly. An alternative approach is to jointly predict 2D and 3D joint positions~\cite{tome2017lifting,VNect_SIGGRAPH2017}. 
%
Many of these approaches, however, do not specifically tackle the occlusion problem. The few existing works that focus on predicting pose in the presence of occlusions either consider only 2D pose~\cite{Fu_2015_ICCV}, or represent 3D human pose as sparse linear combinations of known 3D poses and recover these blend weights from partially observed data~\cite{Huang:2009}. Such approaches, however, are limited to image input, do not consider the temporal dynamics of the human motion, and do not reason explicitly about occlusion arising due to human-object interactions. In contrast, given an initial pose estimate 
(acquired by off-the-shelf 3D pose estimators that do not consider global positioning of the detected skeleton), 
%(in our setup, acquired by the method of Tome et al.~\shortcite{tome2017lifting} who do not consider global positioning of the detected skeleton), 
\name\ jointly reasons about scene geometry and human pose to synthesize both plausible scenes and human motion even in case of moderate to heavy occlusions. In Section~\ref{sec:evaluation}, we show comparisons of the 3D human motion recovered by our method to the recent 3D pose estimation methods.

%% file: 03_overview.tex
\section{Overview} \label{sec:overview}

The input to our method\footnote{Project code/data will be released for research use.} is a monocular video showing a  person interacting with objects. Our goal is to synthesize human performance that fits this input video and a static object arrangement that is plausible and consistent with the inferred human performance.

When watching performance of a human actor in a scene, there are several cues that one can use to recover plausible explanations for objects and performance sequences, even under partial occlusion. Interactions with objects, for example, give cues about both the objects and the motions that are part of the interaction. Walking gives cues about occluded empty space in a room, while occlusion of joints give cues about the location of objects relative to the actor. Such cues, when combined with prior knowledge of typical human-object interactions, help recover plausible human performance and object arrangements (see  Figure~\ref{fig:imapper_illustration}) as described below.%Using these cues requires the application of prior knowledge to the observations given in the video. To synthesize plausible human performance and objects based on these cues, we proceed as illustrated in Figure~\ref{fig:overview} and as described below.        ``  

To represent prior information (see Section~\ref{sec:scenelets_skeletons}) about typical interactions with objects, we build a dataset using PiGraphs of human-object interactions called \emph{scenelets}. Each scenelet contains a short motion clip and a set of static objects in close proximity to the motion. These scenelets capture relationships between the motions of an actor and object arrangements.
Additionally, we get prior information about typical skeleton poses from a selection of pre-trained models of human poses~\cite{tome2017lifting,RogezWS18}.

We then synthesize an output by fitting selection of these models to each part of the video, optimizing an energy function (see Section~\ref{sec:fitting_energy}) that quantifies the consistency of the fitted models with the video and with each other. By consistency, we measure how well the fitted models explain the presence or absence of skeleton joint detections in each video frame, and on other plausibility criteria, such as path smoothness and intersection avoidance. 
We formulate the problem as a global optimization (see Section~\ref{sec:fitting_scenelets_skeletons}). 
%After describing the scenelets (Section~\ref{sec:scenelets_skeletons}), we detail the model fitting process (Section~\ref{sec:fitting_scenelets_skeletons}) then describe the optimization formulation (Section~\ref{sec:fitting_energy}).

We decompose the optimization into several simpler sub-problems to obtain a robust initialization before the final optimization. We start by fitting scenelets in our dataset to a set of regularly spaced time intervals in the video. Fitting is done independently for each scenelet to avoid a combinatorial explosion of possible scenelet combinations. We continue to fit skeletons to all frames not covered by scenelets. This initialization provides sufficient information to commit to a subset of scenelets that constitute our synthesized scene. Finally, we optimize the placement of all chosen scenelets and skeletons with the full fitting energy, including interactions between all fitted models (see  Section~\ref{sec:scene_synth}).

%% file: 04_method.tex
\section{Human Poses and Scenelet Priors }\label{sec:scenelets_skeletons}

\revision{
In occluded parts of a video sequence, \name recovers information about objects and actor motions by fitting interaction models to the video sequence. In unoccluded sequences, we fit static skeleton poses instead. The models we fit to the video sequence represent our prior information about valid human poses and interactions. Next, we will describe these models.
}
%We represent prior information about valid human poses and interactions with two models, described next.

\subsection{Human Pose Prior}
The space of valid static poses is defined with the 3D skeleton pose model described by Tome et al.~\shortcite{tome2017lifting} or Rogez et al.~\shortcite{RogezWS18}. Such models are trained on a large dataset of poses, carefully removing factors such as rotation in the ground plane and left-right \mbox{symmetry}. The authors also take care to capture less frequently occurring poses, giving this model a good coverage of the valid human pose space. Please refer to the original papers for  details. %\hl{\cma{TODO: mention \lcrnet.}}

\subsection{Scenelet Prior}
We use the PiGraphs dataset~\cite{savva2016pigraphs} to model interactions between humans and objects. The original dataset contains a set of scenes with a commodity depth sensor, each containing a human performance and a set of labeled objects captured. Labels group objects into a small set of categories, such as tables, sofas, chairs, and bookshelves. From this dataset, we extract short sequences $\mathcal{L}_1, \dots, \mathcal{L}_m$ showing interactions between the actor and objects. We call such short sequences \emph{scenelets}.
%
%\mypara{Scenelets}
%In addition to this input, we have access to prior knowledge about typical interactions between actors and objects, in a dataset of scenelets $\mathcal{L}_1, \dots, \mathcal{L}_m$.

Each scenelet consists of a short motion clip with known 3D joint positions and a set of static objects. We denote with $s^{lt}_k$ the location of skeleton joint $k$ in frame $t$ of scenelet $l$. Objects $O^l = \{o^l_1 \dots o^l_n\}$, of scenelet $l$ are defined by a placement $p$, a rough approximation of their geometry $\kappa$, and a label $b$; \ie each object is encoded as triplet $o = \{p, \kappa, b\}$. We assume objects can only rotate around the up direction, giving us four degrees of freedom for the placement of an object: $p = (x, y, z, \theta)$, where $x$, $y$, and $z$ are the location, and $\theta$ the orientation of the object. Similar to the original dataset, we approximate geometry of objects by unions of cuboids; and the label $b^l_i$ describes the object type from the predefined set of categories.
%$K^l_i = \bigcup_j \kappa^l_{i,j}$ $\kappa^l_{i,j} = (\mathbf{c},\mathbf{s},\mathbf{\Theta})$ is cuboid $j$ of this object, defined by a center $\mathbf{c}$, a size $\mathbf{s}$, and a 3D orientation $\mathbf{\Theta}$.
%Each object additionally has a label $b^l_i$ describing its type.
Both motion clip and objects are stored in the local coordinate frame of the scenelet, defined by the pelvis location and the forward-facing direction of the skeleton in the center frame of the motion clip.
%\duygu{shall we denote the scenelets with $\mathcal{L}_1 \dots \mathcal{L}_m$ since later we use $l$ to refer to them?}

\mypara{Scenelet parameterization.} When constructing scenelets, we make a design choice regarding the length of the motion clip used for each scenelet based on the following considerations. 

First, the speed at which an interaction is performed should not affect the contents of a scenelet. For example, if a scenelet captures a fast `sitting-down' performance, then a slower version of the same interaction should also be captured by a single scenelet. This property is necessary to ensure that interactions captured by scenelets are comparable.

Second, scenelets should represent interactions that are local in space. Keeping spatial locality simplifies the subsequent optimization, since it reduces the number of potential interactions between scenelets. We have found that using scenelets where the skeleton (\eg the pelvis joint) traverses constant arc length  satisfies both time invariance and locality. To reduce the effect of noise, we compute the arc length on a smoothed %version of the 
pelvis trajectory, using $10$ iterations of a moving average  with an arc length radius of \mbox{$1$ cm}.

%, which we represent with cuboids. $o^l_i, \dots, o^l_{o_l}$. We represent them with cuboids 

\mypara{Scenelet construction.}
The PiGraphs dataset describes human performance with a set of $16$ joint locations per frame. We start by sampling the center of each scenelet's motion clip at regularly spaced intervals on the arc length of the pelvis joint's trajectory. The start and end of the motion clip in each scenelet is then defined as this center point plus/minus half the scenelet length.
The objects of the scenelet are chosen as the subset of objects roughly within arm's reach of the actor at any point in the motion clip, \ie  the objects the actor can potentially interact with. In our test, we pick objects within \mbox{$1$ m} radius when projected to the ground plane.

\mypara{Scenelet descriptors.}
In order to compare two scenelets and compute their distance, we define two descriptors for each scenelet: a \emph{motion descriptor} $\Psi$ and an \emph{object descriptor} $\Phi$.

The {\em motion descriptor} $\Psi$ compactly describes the motion clip of a scenelet with a fixed-length vector. It is the concatenation of a fixed number of static pose descriptors $\Psi := (\psi_1, \dots, \psi_{k})$, sampled evenly over the motion clip. In our experiments, we use $k=15$ samples. Static pose descriptors are based on $14$ robust joint-line distances (see supplemental for details) as suggested in Zhang et al.~\shortcite{zhang16}. Distances between these descriptors are defined using a weighted L$_2$ distance, assigning more weight to center frames.  
The descriptors $\psi_i$ should evenly cover the motion, and similar to the length of a scenelet, they should be invariant to the speed of the motion. We evenly distribute these samples along the trajectory of the motion clip in a $17$D space of the combined pose descriptor and global skeleton location (taken to be the 3D location of the pelvis).
%We use a weighted L2 distance in this pose descriptor space:
%\begin{equation*}
%    d^2(\Psi_1, \Psi_2) = \sum_{t=1}^{15} \|\psi^1_t - \psi^2_t\|^2_2\ \mathcal{G}(t|8,2.33),
%\end{equation*}
%where the weight emphasizes the center frames: it is a Gaussian $\mathcal{G}$ centered at the middle frame (frame 7) and falling off to nearly zero at the first and last frames of the scenelet.

\begin{figure}[h!]
  \includegraphics[width=\columnwidth]{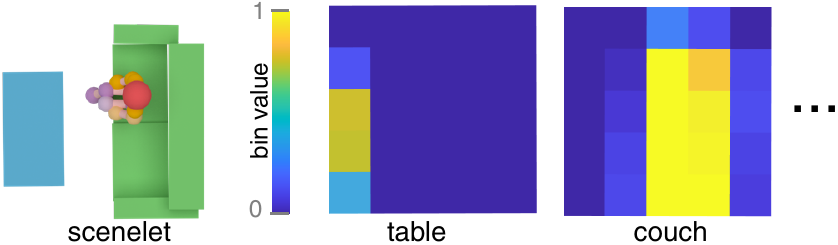}
  \caption{Object descriptors compactly represent the object arrangement of a scenelet. One $5\times 5$ histogram per category stores the layout of objects of this category relative to the scenelet center.}
  \label{fig:obj_descriptors}
\end{figure}

\begin{figure*}[t!]
    \centering
    \includegraphics[width=\textwidth]{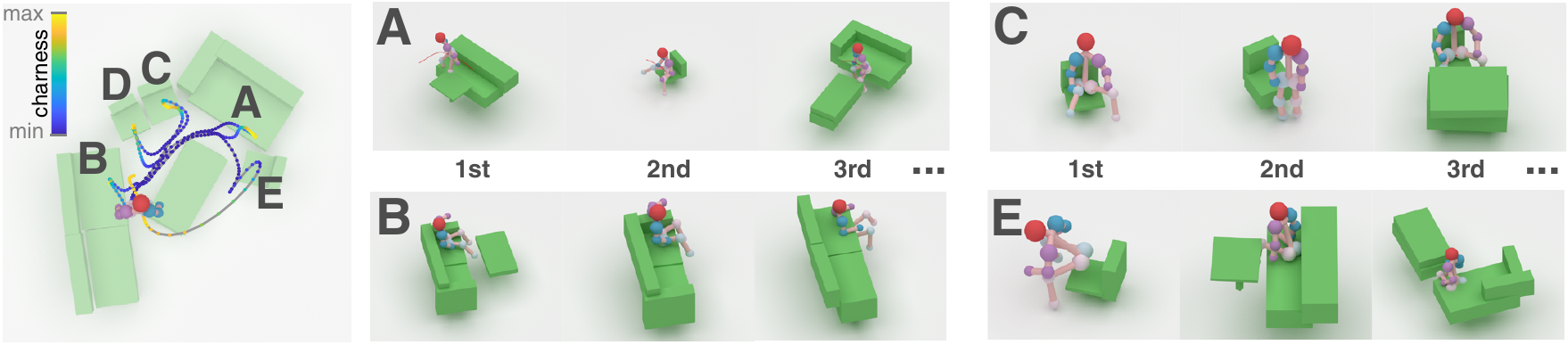}
    \caption{[\Scene{livingroom00}] Fitting scenelets to characteristic sequences. Scenelets are fit to charness maxima in the video sequence. On the left we show the charness of the human's pelvis trajectory in one of our synthesized scenes. The scenelets on the right are the best 3 candidates for an independent fit to each of the gaps, without interaction between scenelets. Note that most of these candidates plausibly fit their video sequence.
    }
    \label{fig:top_view}
\end{figure*}

The {\em object descriptor} $\Phi$ captures the object arrangement as a set of histograms. We store one histogram per object category. The histograms capture the 2D placement of objects, projected to the ground plane. Our histograms are $5\times 5$ square grids (with center poses facing the forward direction) and each bin $\Phi_j$ describes to what extent any object of the same category in the scenelet is located in this bin. We define the value in a bin as the maximum coverage of the bin by any object. To handle both objects smaller and larger than the bin, we normalize by the smaller of either the bin area or the object area:
\begin{equation*}
    {\Phi_j := \max_i \left\{ {A\left(\Lambda(o_i) \cap \phi_j\right)}/{\min\left(A\left(\Lambda(o_i)\right),\ A(\phi_j)\right)} \right\} },
\end{equation*}
where $\Lambda(o_i)$ is the projection of object $o_i$ to the ground plane, $\phi_i$ is the part of the ground plane covered by bin $i$ and $A(x)$ is the area of $x$. Figure~\ref{fig:obj_descriptors} shows an example of an object descriptor.

\begin{figure}[h!]
  \includegraphics[width=\columnwidth]{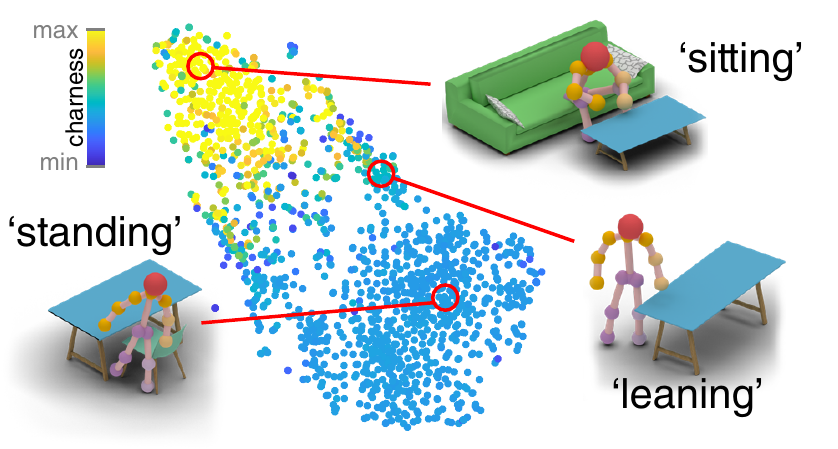}
  \caption{Characteristicness of our scenelet dataset over pose descriptor space. We show a t-SNE embedding with our descriptor distance. Warmer colors denote more characteristic scenelets. The close-ups show scenelets with a standing sequence, a sitting sequence, and a leaning sequence.
  %The characteristicness for `couch' objects is shown below the skeletons.
  Standing sequences have low charness because they are not specific to the neighboring objects.}
  \label{fig:charness}
\end{figure}

\mypara{Charness.}
The characteristicness, or \emph{charness} for short, of a bin in an object descriptor, describes how typical an activation of this bin is for similar motion clips. For example, for a sitting-down motion, a couch or a chair at the center bin of the histogram will have high charness. This charness score helps to distinguish between objects that are related to a given interaction, and objects that are near the motion, but unrelated to the interaction. The charness of an object descriptor bin is computed as a weighted average of that bin's activation over similar motion clips, where similarity is defined with a Gaussian kernel in the space of motion descriptors:
\begin{equation*}
h^l_j := \frac{\sum_{k=1}^m \Phi^k_j\ \mathcal{G}(d(\Psi_k, \Psi_l)| 0, \sigma) p_k^{-1}}{\sum_{k=1}^m \mathcal{G}(d(\Psi_k, \Psi_l)| 0, \sigma) p_k^{-1}},
\end{equation*}
where $h^l_j$ is the characteristicness of bin $j$ in scenelet $l$, $\Phi^l_j$ is the bin value of scenelet $l$, $\mathcal{G}$ is a Gaussian kernel taken over the distance $d$ between the motion clip descriptors defined earlier (we empirically set $\sigma = 13$), and $p_l$ is the \emph{density} of scenelets at the origin of scenelet $l$. We measure the density of scenelets in the space of the original PiGraph scene that scenelet $l$ was obtained from. It is defined as the spatial density of the origins of all scenelets that were obtained from this scene. The intuition behind dividing by this density is to remove the bias we would otherwise introduce due to multiple scenelets taken from nearby parts of the scene. Note that multiple scenelets showing the same part of the same scene are correlated and would bias the charness towards this particular scene arrangement.
Finally, we define the charness of a scenelet as the maximum bin characteristicness:
\begin{equation*}
    H^l := \max_j(h^l_j).
\end{equation*}
Scenelets with high charness are likely to contain interactions with objects, since they have objects within interaction range that are typical for the scenelet's motion. We will use the charness of a scenelet to determine which sequences of a video are likely to contain interactions between the actor and objects (see Figure~\ref{fig:charness}).

\section{Fitting Scenelets and Skeletons}
\label{sec:fitting_scenelets_skeletons}
\revision{
The pose space and the scenelets discussed in the previous section give us strong priors for human poses, and interactions with objects, respectively. In occluded parts of a video sequence, we can fit these models to the sparse set of observations, giving us likely explanations for the parts that could not directly be observed. In this section we describe this fitting problem as a search in a high-dimensional parameter space, and in the next section we define an energy in this space that we minimize to get an optimal fit.
}

In each frame of the input video, we detect 2D image-space skeleton of the actor, consisting of $n_j$ joint locations using an off-the-shelf pose detector.  \revision{In our experiments we use either (i)~2D keypoints detected by CPM~\cite{wei2016} and grouped based on the grouping heuristic of Tome et al.~\shortcite{tome2017lifting} or (ii)~\lcrnet~\cite{RogezWS18}.} %or OpenPose \cma{(we have no scenes with OpenPose2D+Tome3D, but could have)}} \sout{Tome et al.~\shortcite{tome2017lifting}}. 
Given a video with $n_\nu$ frames, for each joint $k$ detected in frame $t$, $u^t_k \in \mathbb{R}^2$ denotes its location and with confidence $c^t_k \in [0,1]$. \revision{While confidences are directly provided by the method of CPM, when using \lcrnet we compute them based on the per-joint pose proposal variance (see Appendix~\ref{appendix:confidence} for detail).}

Our goal is to synthesize a scene consisting of 3D joint locations $q^t_k~\in~\mathbb{R}^3$ for each video frame, describing the human performance, and a set of objects $O = \{o_1, \dots o_{n_o}\}$. %Each object is defined by a placement and an index into a dataset of objects. We assume objects can only rotate around the up direction, giving us four degrees of freedom for the placement of an object: $p^o = (x, y, z, \theta)$, where $x$, $y$, and $z$ are the location, and $\theta$ the orientation of the object.
The 3D joint locations at each frame are obtained by fitting either a scenelet or a static skeleton to the 2D joint detections in the video, and objects are taken from the fitted scenelets.
%
%\mypara{Choosing scenelets or static skeletons}
%For each frame of the video, we have a choice between fitting a static skeleton, or one of the scenelets in our dataset.
%
Which of the two models we fit to a given part of the video depends on two factors: the estimated amount of joint occlusion observed in the video (\ie the confidence of the joint detection signal) and the estimated probability of object interactions.

%
%In the following, we describe the fitting parameters of scenelets and static skeletons.

%We regularize our joints and objects by fitting either scenelets to short sequences of the video or 3D skeletons to single time steps, where the skeletons are taken from the space of valid skeleton poses defined in Tome et. al.~\cite{tome2017lifting}. Which of the two models we fit to a given part of the video depends on two factors: the amount of joint occlusion observed in the video (i.e. the strength of the joint signal) and the presence of object interactions. The latter can be estimated based on the charness of scenelets fitted to the video part.
%
%When fitting scenelets and 3D skeletons to the 2D joint detections in the video, our goal is to obtain 3D joint locations $q^t_k \in \mathbb{R}^3$ for each video frame, describing the human performance, and a set of static objects $o_1 \dots o_{n_o}$.

\mypara{(i) Fitting scenelets.}
Video sequences that contain interactions with objects or occluded human performance, are modeled with scenelets allowing allow us to both populate the scene with objects involved in interactions and explain occlusions of joints due to these objects. Thus, joint occlusions can help in both choosing and placing the scenelets: for a given video sequence, we would like to choose and place a scenelet such that the scenelet objects explain the joint occlusions observed in the video sequence.

%Each frame in the video can be the starting point of no more than one scenelet. 

We start by modeling the assignment of scenelets in our dataset to time intervals in the video.  Given a video with $n_\nu$ frames and a dataset with $m$ scenelets, only a single scenelet can start at any frame of the video. The scenelet assignment can therefore be expressed with a binary matrix $\mathcal{X} \in \{0,1\}^{m\ \times\ n_\nu}$
\begin{equation*}
\mathcal{X}_{lt} =
\begin{cases}
1 & \text{if scenelet $l$ starts at video frame $t$} \\
0 & \text{otherwise}.
\end{cases}
\end{equation*}
The constraint that scenelets should not overlap in time can then be formulated as
\begin{equation*}
\eta_t = \sum_{l} \sum_{i=1}^{\max(t, n_l)} \mathcal{X}_{l(1+t-i)} \leq 1,\ \ t=1\dots n_\nu,
\end{equation*}
where $\eta_t$ is the number of scenelets assigned to frame $t$, and $n_l$ is the number of frames in scenelet $l$.
%Unlike for 3D skeletons, for scenelets we optimize for both placement \emph{and} choice. 
%Unlike for 3D skeletons, for scenelets we can fit any frame of any scenelet in our dataset to a given frame $t$.
%
%When fitting frame $i$ of a scenelet to a given frame $t$, the scenelet starting point is 
%
%Since only a single scenelet can start at any frame $t_s$ of the video, we can define a single transformation per video frame.
%
%We denote the placement of the scenelet starting at frame $t_s$ as $p^s_{t_s} = (x, y, z, \theta)$.
%
%The joint location $(\check{q}^t_k)^l_i$ for video frame $t$ obtained from frame $i$ of scenelet $l$ can then be defined as:
%
%At each frame of the video, we can choose up to one scenelet. Given a video with $n_\nu$ frames and a dataset with $m$ scenelets, the choice of scenelets can therefore be expressed with a binary matrix $\mathcal{X} \in \{0,1\}^{m\ \times\ n_\nu}$, where $\mathcal{X}_{lt}$ is $1$ if scenelet $l$ of the dataset starts at frame $t$ of the video, and $0$ otherwise. Columns of this matrix must sum up to no more than $1$.
%
Since only a single scenelet can start at any frame $t$, we model scenelet placement with one set of parameters per frame $P = \{P_1 \dots P_{n_v}\}$, where $P_t = (x, y, z, \theta)$ is the placement of the scenelet starting at $t$, with $x$, $y$, and $z$ the location and $\theta$ the orientation of the scenelet.
The 3D joint locations $\hat{q}^t_k$ in video sequences covered by scenelets can then be defined as a function of the placement $P$ and the scenelet assignment $\mathcal{X}$:
\begin{equation}
\label{eq:fitting_scenelets}
\widehat{q}^t_k(P, \mathcal{X}) := \sum_{l} \sum_{i=1}^{\max(t, n_l)} \mathcal{X}_{l(1+t-i)}\ T(P_{(1+t-i)}) s^{li}_k,
\end{equation}
%(\check{q}^t_k)^l_i = T(p^s_{(t-i)}) s^{li}_k,
where $s^{li}_k$ is the 3D position of joint $k$ in frame $i$ of scenelet $l$, and $T(P_t)$ is the transformation to placement $P_t$.

Finally, the objects in the scene are obtained from all scenelets that have been assigned to the scene: 
\begin{equation*}
    O(P, \mathcal{X}) := \bigcup_{\{(l,t)\ |\ \mathcal{X}_{lt} = 1\}} T(P_t, O^l),
\end{equation*}
where we denote with $T(P, O)$ the transformation of objects in $O$ to the placement $P$, \ie $T(P, O^l) = \{(T(p),\kappa,b)\ |\ (p,\kappa,b) \in O^l\}$.
%\paul{todo: object choice and placement (which ones do we pick if they intersect?)}

\mypara{(ii) Fitting static skeletons.}
%For parts of the video that contain an unoccluded human performance without object interactions, we can use more traditional skeleton pose fitting to obtain synthesized joint positions. Since the degrees of freedom for human poses is smaller than for human-object interactions, the space of possible human poses can be covered more accurately than the space of possible human-object interactions. Thus, fitting 3D skeletons to the video gives us better performance in unoccluded sequences that do not contain interactions.
For parts of the video that contain an unoccluded human performance without object interactions, we can fit static skeletons to each frame. Since the number of degrees of freedom for human poses is smaller than for human-object interactions, the space of possible human poses can be covered more accurately than the space of possible human-object interactions. Thus, fitting static skeletons to the video gives us better performance in unoccluded sequences that do not contain interactions.

\revision{The aforementioned 3D pose reconstruction methods~\cite{tome2017lifting,RogezWS18}} retrieve the best matching 3D skeleton pose for a given frame. This pose is defined in the {\em local} space of the skeleton and does not give us the placement of the skeleton in the scene. We fit the retrieved 3D skeleton to our video by optimizing the 3D placement of the skeleton, using the fitting energy described later in Section~\ref{sec:fitting_energy}.
%
%We extend this method to give use the 3D placement of the skeleton in each frame, by minimizing energy defined in the previous section.

Skeletons are only fitted to frames that do not have any scenelet assignment. The joint locations $\check{q}^t_k$ for video sequences that are not covered by scenelets are then defined as:
\begin{equation}
\label{eq:fitting_skeletons}
    \check{q}^t_k(P,\mathcal{X}) = (1 - \eta_t)\ T(P_t) a^t_k,
\end{equation}
where the first term is only non-zero if no scenelet is assigned to frame $t$, and $a^t_k$ is the local skeleton pose computed by \revision{using Tome et al. or \lcrnet}, $P_t = (x, y, z, \theta)$ is the placement of the skeleton in frame $t$, and $T(P_t)$ is the transformation to placement $P_t$.
%\duygu{todo: we define $q_k^t$ both as a variable and a function, would be good to change this notation}
%Due to our flat scene assumption, we can omit the placement height.
%We can then optimize the energy defined in Equation~\ref{eq:energy} over $p^a_t$ instead of $q^t_k$ to find the optimal skeleton placements for any video sequence.
%
Combining Equations~\ref{eq:fitting_scenelets} and~\ref{eq:fitting_skeletons}, we  define the location of any joint $q^t_k$ in the video as:
\begin{equation}
q^t_k(P,\mathcal{X}) = \widehat{q}^t_k(P,\mathcal{X}) + \check{q}^t_k(P,\mathcal{X}).
\end{equation}
In the following, we will omit the explicit dependence of $q^t_k(P,\mathcal{X})$ and $o_i(P,\mathcal{X})$ on $P$ and $\mathcal{X}$ for a less cluttered notation.

%\paul{todo: model scenelet vs skeleton choice as skeleton only if no scenelet is placed at a frame. And add term to optimization that favours charness, to prefer placing scenelets over skeletons where characteristic scenelets exist for a frame.}
 
%Optimizing the energy defined in Equation~\ref{eq:energy} over $\mathcal{X}$ and $p^s_t$ instead of $q^t_k$ gives us our scenelet placements.

\section{Fitting Energy}
\label{sec:fitting_energy}
\revision{
We have now set up our search space over possible configurations of objects and actor motions, parameterized through the scenelet and pose placements $P$ and the assignment matrix $\mathcal{X}$.
Next, we define an energy in this space that can be minimized to obtain a plausible configuration of objects and actor motions given the observations in the video.
}
We quantify the quality of a given fit as consistency of the fitted models with the video and consistency between the fitted models. Thus, we define an energy function that penalizes inconsistency:
%
% which we measure by our interaction prior. Furthermore, the synthesized human motion and the scene need to be consistent with the input video and each other, which we express with the following energy function:
\begin{equation}
\label{eq:energy}
\argmin_{P, \mathcal{X}} L := w_r L_r + w_o L_o + w_s L_s + w_c L_c + w_m L_m, 
% \label{eq:synt}
\end{equation}
where $L_r$ is the {\em reprojection} error measuring the difference between the 2D joints and the projection of 3D joint locations, $L_o$ penalizes the presence of {\em occlusions} of skeleton joints in the video  that are not explained by occlusions in the synthesized scene, $L_s$ encourages {\em smoothness} among human performance, $L_c$ penalizes {\em intersections} between objects, and $L_m$ penalizes {\em intersections} between the motion clip and objects. Our goal is to synthesize joint locations $q^t_k$ and objects $o_i$ by minimizing this energy over placements $P$ and assignments $\mathcal{X}$ of all fitted models, while 
%$Q_\nu = \{q^t_k\ |\ t=1 .. n_\nu,\ k=1..n_j\}$, and objects $O_\nu = \{o_1 .. o_{n_o}\}$, that minimize this energy while
ensuring a valid human performance. We next describe each energy term in detail.

\mypara{Reprojection term ($L_r$)}  penalizes the distance from the 2D joint locations detected in the video to the corresponding output joints $q$ projected to screen space as commonly defined residue:
\begin{equation}
\label{eq:energy_repro}
L_r = \sum_{t} \sum_{k} \left\|\Pi q^t_k - u^t_k \right\|_2^2 c^t_k,
\end{equation}
%v(q^t_k, O)\
where $\Pi$ is the camera projection matrix, $u^t_k$ are our input 2D joint locations, and $c^t_k$ is the confidence of each detection.
%$v(q^t_k, O)$ is an indicator function denoting if the $q^t_k$ is visible given the objects $O = \{o_1 \dots o_{n_o}\}$ in the scene.

\begin{figure*}[t!]
    \includegraphics[width=\textwidth]{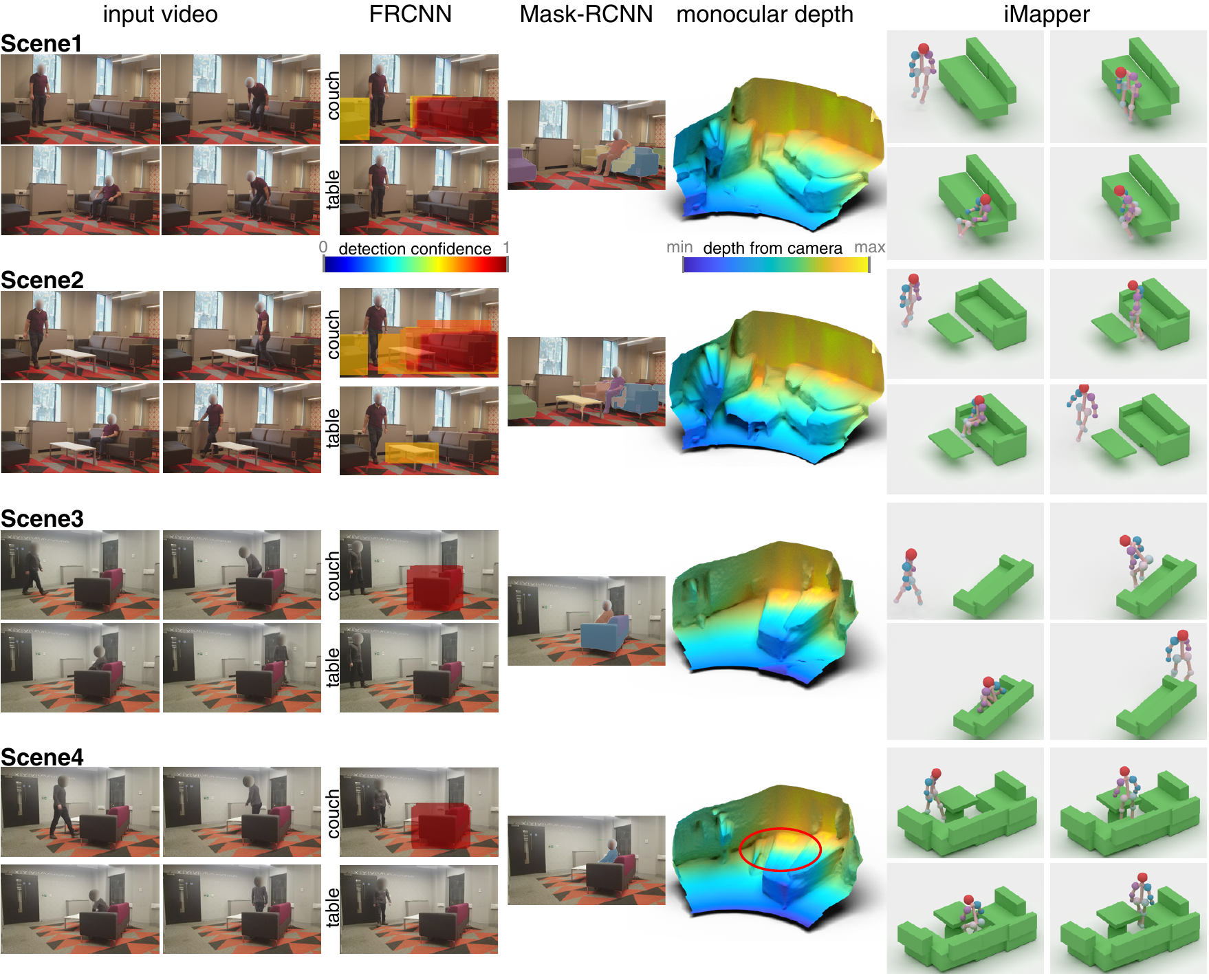}
    %\begin{overpic}[width=\textwidth]{images/qualitativeA.pdf}
    %  \Large
    %  \put(6, 70.5){\color{white}\textbf{\Scene{lobby11}}}
    %  \put(6, 50.75){\color{white}\textbf{\Scene{lobby12}}}
    %  \put(6, 32){\color{white}\textbf{\Scene{lobby18-1}}}
	%  \put(6, 12){\color{white}\textbf{\Scene{lobby19-3}}}
    %\end{overpic}
	%\begin{overpic}[width=0.24\textwidth]{images/mask-rcnn/lobby11-00040.png}
	%	\put(0,0){\color{white}\mbox{\maskrcnn \protect{\cite{He:2017:ICCV}}}}
	%\end{overpic}
	%\includegraphics[width=0.24\textwidth]{images/mask-rcnn/lobby12-00075.png}
	%\includegraphics[width=0.24\textwidth]{images/mask-rcnn/lobby18-1-00060.png}
	%\includegraphics[width=0.24\textwidth]{images/mask-rcnn/lobby19-3-00057.png}
    \caption{Qualitative comparisons to state-of-the-art object detection methods. Note that FRCNN and Mask-RCNN both produce only 2D image-space segments. \name synthesizes scenes that plausibly match the video, even in sequences where objects and the human performance are occluded. \revision{Note, the table is not detected by \maskrcnn in \Scene{lobby19-3}. Encircled in red in the bottom depth map are the typical errors near depth discontinuities obtained from image depth estimators. } 
    }
    \label{fig:qualitativeObject}
\end{figure*}

\begin{figure*}[t!]
	\centering
	\begin{overpic}[width=\textwidth]{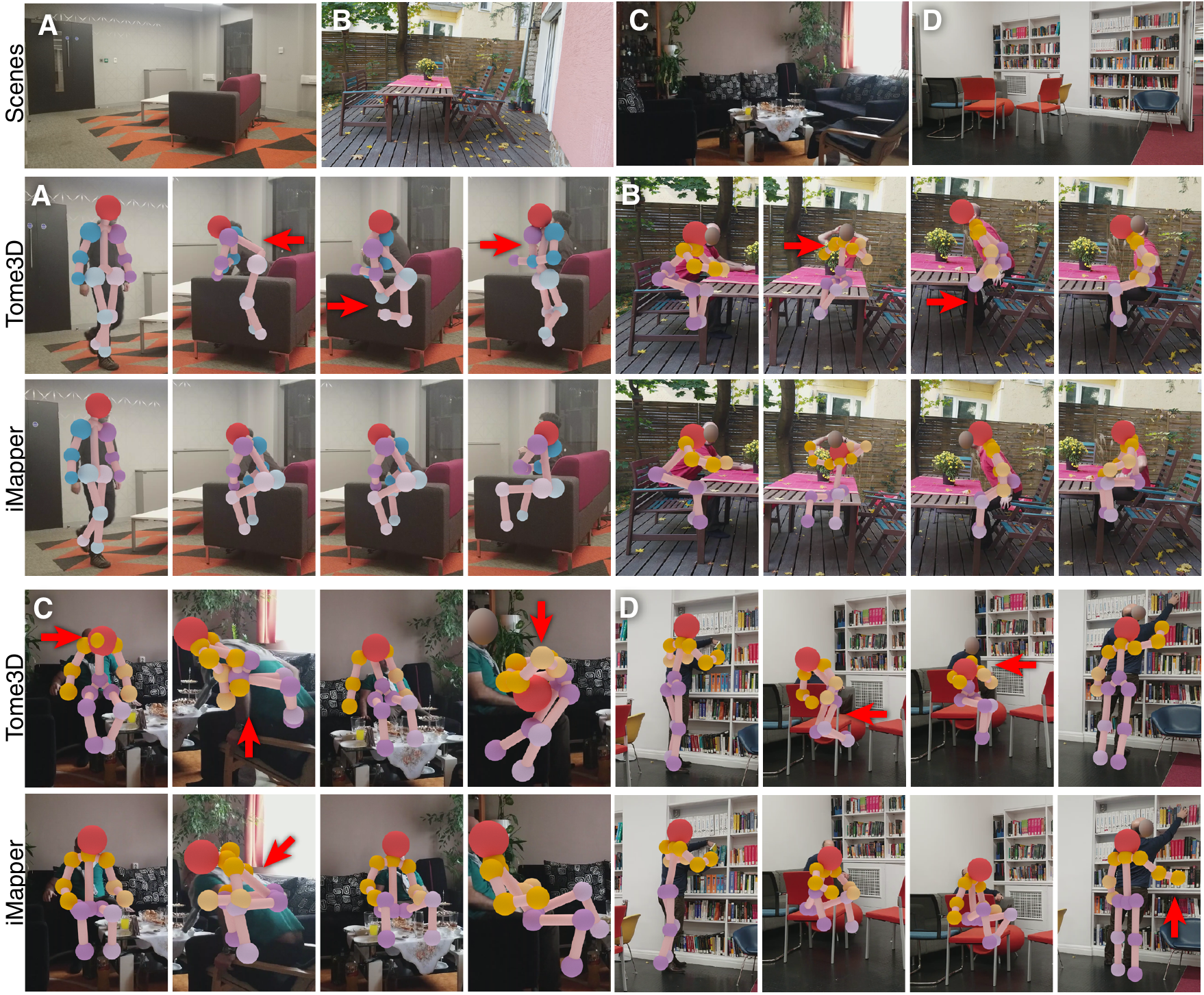}
		\Large
		\put(18,69){\color{white}\textbf{\Scene{lobby19-3}}}
		\put(42,69){\color{white}\textbf{\Scene{garden1}}}
		\put(62.5,69){\color{white}\textbf{\Scene{livingroom00}}}
		\put(90,69){\color{white}\textbf{\Scene{library3}}}
	\end{overpic}
    \caption{Qualitative comparisons to Tome et al.~\shortcite{tome2017lifting}, which produces image-space and local 3D poses. We compare the reconstructed 3D skeleton on four input videos, shown in the top row. Note that Tome and colleagues do not compute world space positions of the skeletons. For better comparison, we position them in world coordinates using the hip locations as estimated by \name. Relevant differences between the methods are marked with red arrows. Note that our method gives plausible skeleton poses in many cases where the method of Tome et al. fails, especially in occluded areas.}
\label{fig:qualitativePose}
\end{figure*}

\begin{figure*}[h]
	\includegraphics[width=\textwidth]{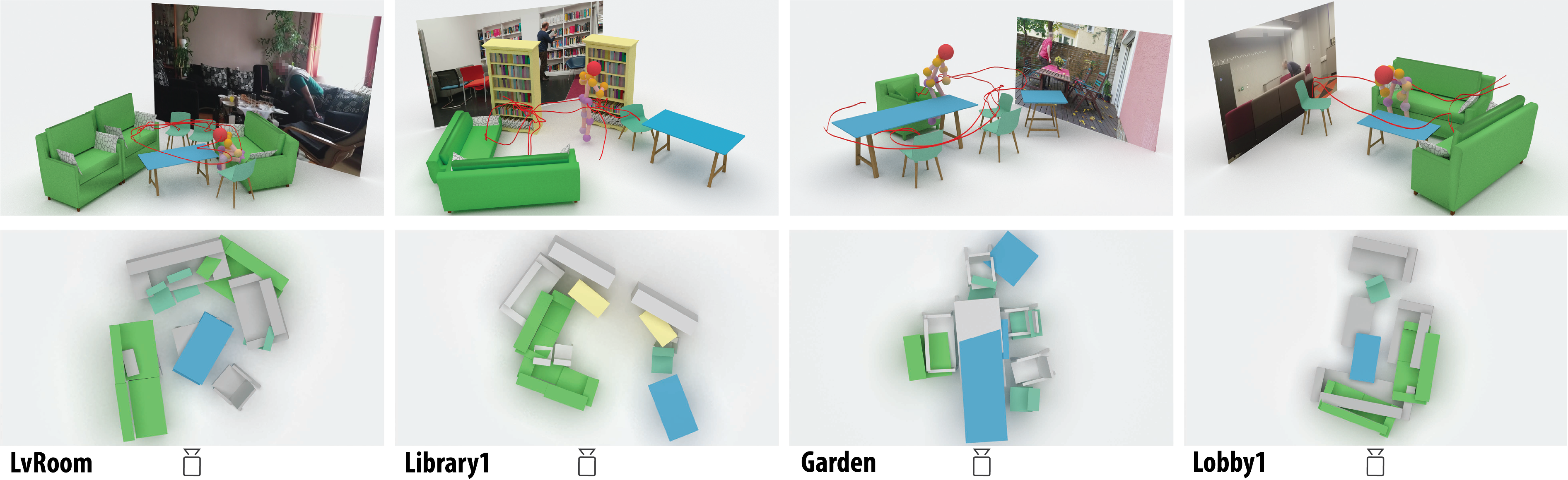}
	\caption{(Top-row)~Plausible object layout and human movement as predicted by \name on various monocular videos. (Bottom-row)~For qualitative evaluation, we overlay, shown from top-view, estimated scene layout versus annotated groundtruth. 
	For quantitative evaluation, please refer to Tables~\ref{tab:occlusions} and \ref{tab:poses}. Please refer to supplmental for videos and results. 
	}
	\label{f:eval_qualitative}
\end{figure*}

\mypara{Occlusion term ($L_o$)} 
 enforces consistency between joint occlusions observed in the video and occlusions of joints induced by synthesized scene objects. Thus, we require the synthesized objects to explain observed occlusions.
%the visibility of the joints induced by the synthesized objects and human motion to be consistent with the 2D joint detections in the input video. %This error discourages synthesized scenes that do not explain the presence or absence of joint detections in the video. 
We assume the person stays in the camera frame for the entire duration of the video. Hence, missing joint detections occur either due to false negatives in the detection method, or due to occluding objects. The reverse is, however, \textit{not} true: the joint detector may, in some cases, also predict the position of occluded joints with high confidence. Therefore, we define an asymmetric occlusion error:
\begin{equation}
\label{eq:energy_occlusion}
L_o = \sum_{t} \sum_{k} F(v(q^t_k, O), c^t_k),
\end{equation}
where $v(q^t_k, O)$ denotes the visibility of joint $q^t_k$ given the scene objects $O$. We obtain non-zero gradients that are necessary for our gradient-based solver by defining $v$ as the signed distance of joint $q^t_k$ to the \emph{occlusion volume} induced by $O$, which is the volume that is not visible from the camera. The asymmetric occlusion error, $F$, for a joint and a set of objects is then defined as:
\begin{equation*}
F(v, c) = 
\begin{cases}
(c-0.5)^2 v^2 & \text{if}\ c-0.5<0\ \text{and}\ v>0 \\
0 & \text{otherwise},
\end{cases}
\end{equation*}
where $c \in [0,1]$. Note that this function is non-zero \textit{only} when low-confidence joint detections are explained by visible joints.

\mypara{Smoothness term ($L_s$)}  ensures continuity of the synthesized motion by measuring the time-derivative of the synthesized joint locations. We approximate this derivative with finite differences:
\begin{equation}
\label{eq:energy_smooth}
L_s = \sum_{t} \left\|q^t_\lambda - q^{t-1}_\lambda\right\|_2^2,
\end{equation}
%\frac{1}{\tau_t - \tau_{t-1}}
where $\lambda$ is the index of the pelvis joint at video time of frame $t$.  

\mypara{Object intersection term ($L_c$)} 
discourages object-object penetration. 
In our flat scene assumption, all objects are placed on the ground plane. We approximate intersections in 2D, using the projections of objects to the ground plane. To obtain non-zero gradients, which are necessary to resolve intersections in a gradient-based solver, we quantify the amount of penetration using signed distance functions:
\begin{equation}
\label{eq:energy_intersect}
L_c = - \sum_{b_i \neq b_j \land \theta_i \neq \theta_j} \left(\int_{\Lambda(o_i)} \delta^-_{o_j}(x)\ dx + \int_{\Lambda(o_j)} \delta^-_{o_i}(x)\ dx\right),
\end{equation}
where $\delta^-_{o_i}$ is the negative part  of the signed distance function of object $o_i$, $\Lambda(o_i)$ is the projection of object $o_i$ to the ground plane, $x$ is a point on the ground plane, $b_i$ is the label of object $o_i$, and $\theta_i$ its orientation. We do not penalize intersecting objects that have the same label and orientation, since we assume these to be representations of the same object placed by different scenelets. For example, a scene with two couches facing one table may be constructed by two scenelets, each placing a couch and the same table. We identify objects to be compatible if they have the same label and orientation.

\mypara{Motion intersection term ($L_m$)}  discourages humans going through objects.  The trajectory of the human motion provides information about empty regions in the scene, since objects may not intersect the motion trajectory. For efficiency, we compute the intersection in 2D on the ground plane and focus on three joints only: the pelvis joint and the two knee joints. In practice, we have found that taking the maximum 2D distance of these three joints to objects allows a reasonable estimation of full 3D intersections in our scenes, since sitting motions can be handled correctly. We use, 
\begin{equation*}
    L_m = \sum_t \max_{q \in \{q^t_\lambda,\ q^t_{\Gamma l},\ q^t_{\Gamma r}\}} \min_i \delta_{o_i}(q),
\end{equation*}
where $\delta_{o}$ is the signed distance function of object $o$, and $q^t_{\Gamma l}$, $q^t_{\Gamma r}$ are the left- and right-knee joints, respectively.

\vspace{\myparaskip}
Note that directly optimizing Equation~\ref{eq:energy} over placements of objects $o_i$ and joint positions $q^t_k$ would neither guarantee the realism of the synthesized human performance, nor its compatibility with the synthesized objects. However, our approach to minimizing this energy by fitting scenelets and static skeletons, as described in the previous section, introduces a good initialization of valid human performances and object interactions that favours realistic scenes and compatibility between objects and human performance.
We obtain joint locations $q^t_k$ and the placements of objects $o_i$ from fitted scenelets and skeletons, and our goal is to optimize the fitting energy over the placements $P$ of skeletons and scenelets (see Equations~\ref{eq:fitting_scenelets} and \ref{eq:fitting_skeletons}) as well as the assignments $\mathcal{X}$.

%We employ two strategies to introduce a prior of valid human performances and object interactions.
%For video sequences with occluded joints or sequences that contain human-object interactions, we obtain 3D joints and object locations by fitting scenelets to the 2D detections in the video. For unoccluded joints in the video that correspond to human performance without object interactions, we instead fit 3D skeletons from a space of valid skeleton poses, extending the method of Tome et. al.~\cite{tome2017lifting}. Both strategies aim at minimizing the energy defined above and will be discussed next.

%
%
%
% \subsection{Initial world-space motion from monocular RGB video}\label{sec:method:init}
\section{Scene Synthesis}\label{sec:scene_synth}

\revision{
Minimizing the energy $L$ gives use a plausible set of objects and actor motions. However,}
due to a difficult parameter domain including integer parameters for scenelet choice $\mathcal{X}$ and a highly non-linear energy function, it is not feasible to minimize the full energy over all parameters in a single optimization. Instead, we approximate the solution by decomposing the optimization. We start with a large set of candidates and evaluate a selection of computationally-efficient energy terms. More complex terms are added in stages, where each stage allows us to filter out low-scoring scenelet candidates. In each stage, we perform an optimization over the placement parameters $P$. The scenelet assignment $\mathcal{X}$ is optimized indirectly by filtering out candidates in each stage of the decomposed optimization instead of directly invoking an integer program. 

\mypara{Static skeleton fitting.} Initially, we do not know if any given sequence of the video contains interactions (since object selection and placements are unknown). Therefore, we start by fitting static skeletons to all frames of the video, \ie we initialize $\mathcal{X}$ to zeros. We optimize for the skeleton placements $P_t$ in each frame $t$. Note that this initial scene does not yet contain objects, so only the reprojection and smoothness terms need to be minimized.

\revision{Depending on the method used to generate input pose detections, we may or may not get valid 3D skeleton guesses for occluded frames. For example, the local 3D pose detector of Tome et al.~\shortcite{tome2017lifting} returns highly unlikely poses in occluded regions, such as poses having their knees above their head. We conservatively discard such estimates and label the corresponding frames as occluded. In addition, we perform outlier detection on the 2D keypoint estimates and define frames as occluded based on a $5\%$ Winsorization of the joint velocities between frames. For such occluded frames, we initially interpolate the joints linearly from unoccluded frames. \lcrnet provides 3D joint guesses even for such frames, thus we keep these estimates even though they are unrealistic (e.g., ankles detected below the floor level). Later, these interpolated or unreliable pose estimates will be replaced by scenelets.}
%\hl{In the case of Tome et al.~\shortcite{tome2017lifting}}, for occluded frames, our static skeletons are unreliable, therefore we initially interpolate the joints in occluded frames linearly from unoccluded frames. Later, these occluded parts will be replaced by scenelets.
%
%A frame is defined as occluded based on two criteria: (i)~We perform outlier detection on the 2D keypoint detections and define frames as occluded based on a $5\%$ Winsorization of the joint velocities between frames. This removes large jumps between frames that are typically caused by a temporary loss of tracking.
%(ii)~We conservatively remove highly unlikely poses returned by the local 3D pose detector of Tome et al.~\shortcite{tome2017lifting}, such as poses having their knees above their head. \cma{This paragraph is true for Denis only, \lcrnet\ has similarly unreliable skeletons, but at least gives a guess}.
%\hl{In the case of \lcrnet\ \cite{RogezWS18}, since 2D and 3D are coupled, 2D features are less reliable in occlusion, and often yields standing poses with ankle detections below floor level. The upper body detections yield an invaluable signal for scenelet fitting.}

\mypara{Scenelet fitting.} To optimize the scenelet assignment $\mathcal{X}$ and optimize placement parameters $P$ for assigned scenelets, we start by identifying frames of the video that contain interactions. 

We fit the scenelets in our dataset (1500) to each video frame in the regularly spaced subset $t \in T'$ and use the characteristicness of the scenelets weighted by their matching quality
%of the top scenelet matches
to determine the probability of an interaction at time $t$. Since we only want to fit scenelets to parts of the video that are occluded or contain interactions, we perform non-maximum suppression of the charness over the video frames and only keep frames that are at charness maxima and above a minimum charness, in addition to frames without static skeletons.

Scenelets are fitted independently, that is, for each scenelet, we evaluate the energy of a scene containing the previously fitted static skeletons plus the single fitted scenelet. Since we are only interested in evaluating the characteristicness of the motion in the video, we do not include the occlusion term in this optimization.

In addition to the charness, the partial fitting energy obtained in this step provides a lower bound for the full fitting energy. We can therefore discard high-energy scenelets. In our experiments, we keep the top $200$ scenelets for each charness maximum. Figure~\ref{fig:top_view} shows some example fits on different parts of a test sequence.

After this first stage of scenelet fitting, we add the motion intersection term to our energy and re-optimize the reduced set of scenelets candidates. The occlusion term is expensive to compute, we evaluate it once for the fitted scenelet candidates and add it to the energy. Based on this energy, we pick the top $3$ scenelets of each charness maximum.

%Optimizing for occlusions is prohibitively expensive for this large set of candidates. For efficiency, we split the optimization for each scenelet into two steps: An optimization without the occlusion term, followed by an optimization of only the rotation of each scenelet with the occlusion term, where the rotation is evaluated at regularly spaced angles in $[0,360\deg]$.
    
%\paul{option1a:} We re-evaluate the full fitting energy independently for the remaining scenelet candidates and build the scene by adding the top candidate from each remaining video frame to the scene.

%\paul{option1b:} We re-fit the the remaining scenelet candidates independently with the full energy and build the scene by adding the top candidate from each remaining video frame to the scene.
    
%\paul{option2a:} We re-evaluate the full fitting energy independently for the remaining scenelet candidates and iteratively build up the scene by adding one scenelet at a time, starting with the best fitting scenelet. After each iteration, we re-evaluate the energy for the scene, which contains the fitted static skeletons, the previously selected scenelets and the single current candidate.

%\paul{option2b:} We re-fit the the remaining scenelet candidates independently with the full energy and iteratively build up the scene by adding one scenelet at a time, starting with the best fitting scenelet. After each iteration, we re-fit the remaining candidates independently, optimizing a scene which contains the fitted static skeletons, the previously selected scenelets and the single current candidate.

\mypara{Refinement.}
%Due to the large number of candidates and the resulting combinatorial explosion, we did not evaluate all interactions between scenelets in the previous step. 
Having committed to a set a smaller set of candidates, we can optimize placements $P$ of all fitted models in the scene, both static skeletons 
\if0
%\cma{, non-characteristic missing static poses get replaced with the motion clip \emph{only} from the scenelet from the scenelets that cover the gap, picked greedily using the best fitting energy from the initial optimization, } 
\fi
and scenelets, using the full energy term. For simpler scenes, we perform one optimization for all combinations of the remaining candidates. In our experiments, the maximum number of characteristicness maxima was $5$, giving us a maximum of $3^5$ combinations to evaluate. We keep the combination that results in the scene with the lowest fitting energy. \revision{For more complex scenes, \eg multi-person scenes with more ambiguous motion (\eg sitting all the time) we list the top 5 diverse candidates. By diversity we prefer scenes not containing same scenelets in nearby times.}

\mypara{Object selection.}
Finally, we resolve intersections between objects. Recall that the object intersection term does not penalize intersections between compatible objects, \ie  objects with the same orientation and category. For each such intersection, between objects $o_A$  and $o_B$, we remove the object that results in a scene with higher energy. The result of this step is our final scene.

%% file: 060_evaluation.tex
\section{Results and Discussion} \label{sec:evaluation}

\revision{
We tested \name on a range of input monocular videos of varying complexity, including both in-house and out-of-house sequences, such as old movies. Table~\ref{tab:occlusions} shows statistics of these videos.
%$14$ scenes.
%
For in-house sequences, we first describe how groundtruth annotations were created, both for object placements and for actor movement.
Based on this benchmark dataset, we then qualitatively and quantitatively evaluate the performance of \name separately for object placement and actor pose accuracy. We further compare our method against dedicated object detection and pose estimation methods.
For out-of-house sequences, such as old movies, suitable groundtruth for objects and actor poses is not always available, in these cases we only provide qualitative evaluation.
Please refer to the supplemental for the full input videos and ground truth annotations.
}

\input{060_table_performance.inc}

\mypara{Groundtruth dataset.}
\revision{
In order to create a groundtruth benchmark dataset, we annotated both object locations and 3D world-space poses for the in-house video sequences.

For object locations, we physically measured the scene objects' dimensions and positioned objects in the scene to minimize video reprojection error, using known camera intrinsics. We also added labels (\eg. `chair') for each individual object.

For human poses, manually annotating the 3D pose in each frame is not feasible, so we used an assisted approach to generate the groundtruth. We started with estimated 2D joint locations~\cite{wei2016} and then manually corrected them. These corrected 2D locations were then lifted to 3D using the reprojection and smoothness energy terms described in Section~\ref{sec:fitting_energy}. Finally, we inspected the output, manually corrected 3D poses, and added these corrections as additional constraints to the optimization. This process was repeated until we found no more significant errors. The number of corrections depended on the amount of occlusion in the scene.

\mypara{Multi-actor scenes.}
We described our method in the context of single actors. However, for scenes with multiple actors, our method generalizes easily when input 2D tracks come with correspondence information over time. We developed a semi-automatic labeling method that optimizes a Markov Random Field (MRF) energy term to find correspondences between skeleton detections, given some manual guidance (see Appendix~\ref{sec:multi_person_anno} for details). This labeling tool was employed for the following multi-person scenes: \Scene{office2-1}, \Scene{Library2}, \Scene{angrymen00}, and \Scene{Grease}.
}

\subsection{Qualitative Evaluation}

\revision{
First, we show qualitative results of our method on several example scenes. We demonstrate that \name finds plausible object arrangements and actor poses for occluded parts of the scene by rendering our 3D result from a different viewpoint than the source video. Note that in these views, the recreated human motions may appear noisy as our scenelets are based on raw Kinect-based captures included the PiGraph dataset.

Figures~\ref{fig:teaser}, \ref{f:eval_qualitative}, and \ref{f:eval_qualitative2}, we  show results for 9 different scenes (more examples/videos in the supplemental). For each of these scenes, we show a reference video frame in the background on a plane facing the camera. The trajectory of the actor's pelvis is shown as a colored line and the colored skeleton shows an occluded actor pose. Please refer to supplemental for full videos of the reconstructed actor motions. To better visualize the detected orientation of objects, we use object bounding boxes to scale and place category specific proxy object geometries in our scenelets -- please note the meshes are {\em not} output of our method. In the future we plan to update our dataset to use detailed models in the scenelets, making this step unnecessary.

In Figure~\ref{f:eval_qualitative}, we show scenes from two old movies: `12 Angry Men' and `Grease'. To our knowledge there is currently no method available to give a plausible reconstruction of these scenes, due to significant occlusions. Our method finds a plausible object layout and actor motion in these scenes based on the observed interactions. The remaining examples in the figure show results on in-house scenes with varying amounts of occlusion.

Note how much information is encoded in the interactions, enabling us to generate scenes close to the original scenes.
%with minimal human supervision for choosing among candidate scenes.
We can, however, not hallucinate objects that are not interacted with, or obtain information about exact dimensions of objects from interactions.
}

\begin{figure}[b!]
    \centering
    \begin{overpic}[width=\columnwidth]{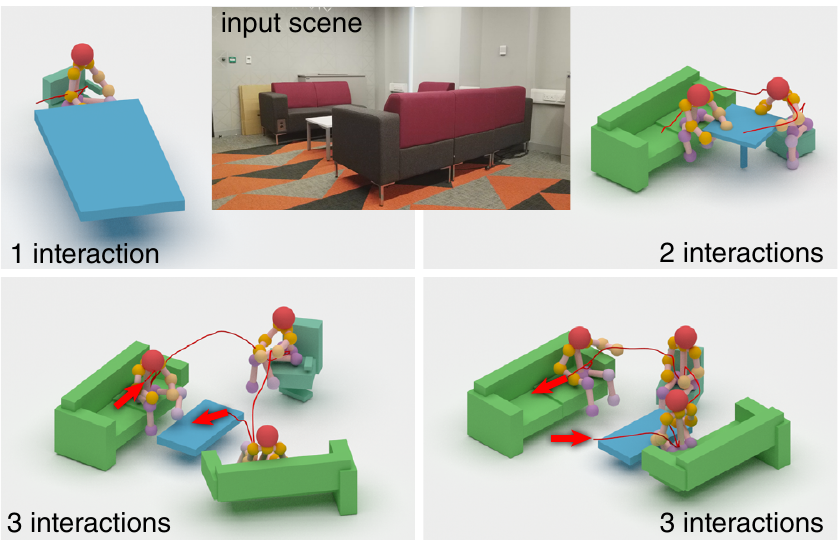}
    	\Large
    	\put(29.5, 32.5){\color{black}\textbf{\Scene{lobby24-3-3}}}
    	\put(51.5, 32.5){\color{black}\textbf{\Scene{lobby24-3-1}}}
    	\put(29.5, 0.5){\color{black}\textbf{\Scene{lobby24-3-2}}}
    	\put(51.5, 0.5){\color{black}\textbf{\Scene{lobby24-2-2}}}
    \end{overpic}
	%\begin{overpic}[width=\columnwidth]{images/mask-rcnn/lobby24-2-2-00095.png}
	%	\Large
	%	\put(2,50){\color{white}\textbf{\maskrcnn} %\protect{\cite{He:2017:ICCV}}}
%	\end{overpic}
    \caption{Scenes can be explored over time. We show results of four different videos taken from the same scene. As interactions with more objects are made available, we can recompute the results to synthesize additional objects. Variations of scene explorations, for example performing the interactions in reverse order, as shown in the bottom row, give slightly different, but comparable and plausible results. }
    \label{fig:scene_evolution}
\end{figure}

\mypara{Scene exploration over time.}
One advantage of \name\ is that the reconstruction quality of the scene improves over time as more interactions `reveal' the true underlying scene. As shown in Figure~\ref{fig:scene_evolution}, as the same environment is explored over time, our system recovers larger parts of the object arrangement.
Further, perturbations to the input (\eg in the form of the same action being performed by different people, or at different times) lead to slightly different, but still plausible and consistent reconstructions of the scene and the interactions.

\subsection{Quantitative and Qualitative Comparisons}\label{ss:comparisons}

\revision{
In the following, we compare both the scene layout and the actor motion recovered by \name to dedicated object detection and pose estimation methods.
}

\mypara{Scene Layout.}
\revision{
We compare against two types of methods: per-frame region detection using \mbox{\frcnn~\cite{Ren:2017:FRT}} and \maskrcnn \cite{He:2017:ICCV}, and per-pixel monocular depth estimation~\cite{ChakrabartiSS16}.

\begin{figure}[t!]
	\includegraphics[width=\columnwidth]{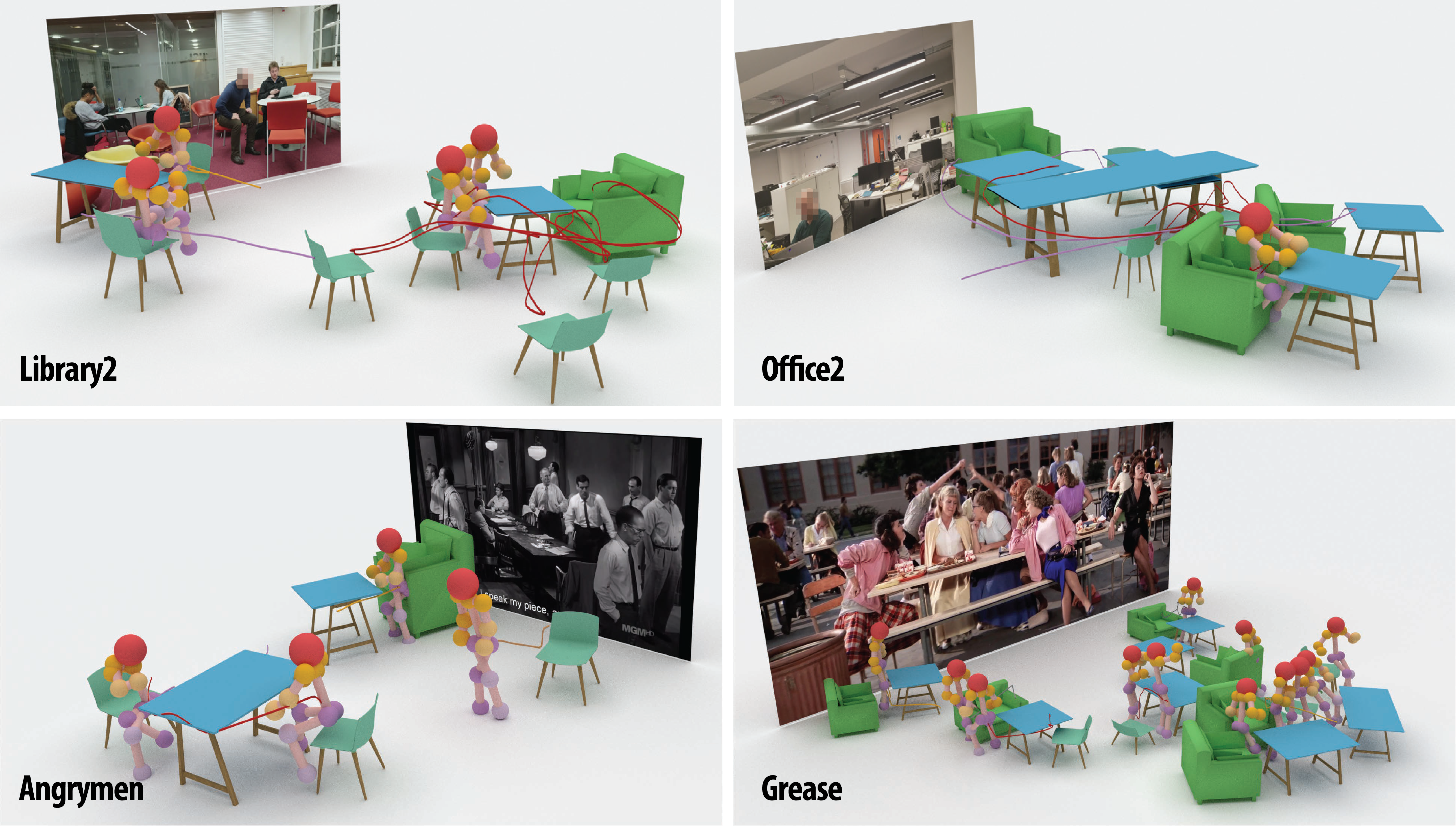}
	\caption{Scene layout and human motion estimated by \name on monocular videos containing multiple actors (see supplementary video). Note that in case of actors not moving sufficiently during the sequence (e.g., \Scene{Angrymen}, \Scene{Grease}), the estimated human `motion' can have drifts. }
	\label{f:eval_qualitative2}
\end{figure}

{\noindent \em Region detection methods:}
For the region detection methods, qualitative comparisons are shown in Figure~\ref{fig:qualitativeObject}, and quantitative results are given in Table~\ref{tab:occlusions}, column `MR'. In Table~\ref{tab:occlusions}, we count the number of objects where at least 50\% of the object's region was detected and correctly labeled on average, over all frames. We provide the count of objects that are participating in at least one interaction. Since \name discovers objects through interactions, this is the set of objects we can potentially detect. \frcnn and Mask-RCNN (MR) are designed to detect visible objects, so they naturally fail to detect any objects `hidden' behind visible objects. For MR this is reflected in the table by a low number of detected objects. Note that other systems that rely on \frcnn\ or MR as their primary building blocks will have similar problems in occluded regions. Our method can more reliably recover these occluded objects if they participate in interactions.}

\revision{
{\noindent \em Depth estimation methods:}
For per-pixel depth estimation~\cite{ChakrabartiSS16} method, as seen in Figure~\ref{fig:qualitativeObject}, fourth column, even for visible regions the estimated depths are smoothed out and fail to capture the object specific layout. \mbox{Table~\ref{tab:occlusions}}, column `MonoD', shows the mean and standard deviation of the distances between the predicted and the ground truth object centroids in the scene. For the monocular depth map, we approximate the object centroid as the mean world position of all samples that are inside the 2D region of an object. Objects without a single visible pixel are ignored. Again, the depth map contains only limited information about partially or fully occluded objects, resulting in large errors. In contrast, \name\ produces plausible objects along with their spatial locations. These locations, in turn, provide better occlusion information for 3D human movement, as evaluated next.}

\input{062_quant_joints.tex}

\mypara{Actor Poses.}
\revision{
Most monocular 3D pose detection methods compute only \emph{local} 3D poses, i.e., joint locations relative to pelvis, limiting our choice of baselines. We compare to Vnect~\cite{VNect_SIGGRAPH2017}, Tome et al.~\shortcite{tome2017lifting}, and  LCRNet++3D~\cite{RogezWS18}. Both Tome et al. and LCRNet++3D output local 3D joint locations, and do not provide world-space coordinates, only VNect also provides world coordinates.
%we provide the methods {\em our estimated pelvis positions} to lift the local predictions to world space and report them in the following comparisons.
For all quantitative comparisons, we report the root mean square error (RMSE) over 14 joint locations. This error is computed in three spaces: 2D image space, (with a resolution of $1920 \times 1080$), local 3D (pelvis-relative) space, and world space, where available.
}
\revision{
{\noindent \em Vnect: }
A qualitative comparison to Vnect is given in Figure~\ref{fig:vnect comparison}. Since we do not have direct access to their source code, we compare to this method on the \emph{MPI-INF-3DHP} dataset~\cite{mono-3dhp2017}, where a relatively high-quality ground truth is available. VNect was designed for unoccluded human motion, so results in occluded areas are not robust. Quantitative results are given in Table~\ref{tab:poses}.
As expected, \name can improve upon Vnect in occluded scenes, resulting in better local and world-space 3D predictions.
%\paul{summarize results}

{\noindent \em Tome et al. and LCRNet++3D: }
Figure~\ref{fig:qualitativePose} shows qualitative comparisons to Tome et al. While non-occluded poses are very close to our method (up to the smoothness term), in occluded frames, Tome3D returns unrealistic poses. \name, however, by fitting scenelets to these occluded sequences returns more realistic (partial) poses that are visually closer to the pose observed in the video. Note that in the last column of {\bf D}, the reprojection is slightly worse for our method, since the scenelet database did not contain a sufficiently close match to describe this kind of motion.

Table~\ref{tab:poses} shows quantitative comparisons to both Tome et al. and LCRNet++3D.
LCR-Net performs well in the less heavily occluded scenes shown near the top of the table, but performance drops in the presence of occlusions. We observe a similar trend for Tome3D, but with a larger error on average. \name can improve upon both methods in more heavily occluded scenes, where we can rely on interaction priors to give us information about hidden joints that the two other methods cannot correctly predict.
%\paul{summarize results}.
}

\subsection{Ablation study}
\begin{figure}
  \begin{overpic}[width=\columnwidth]{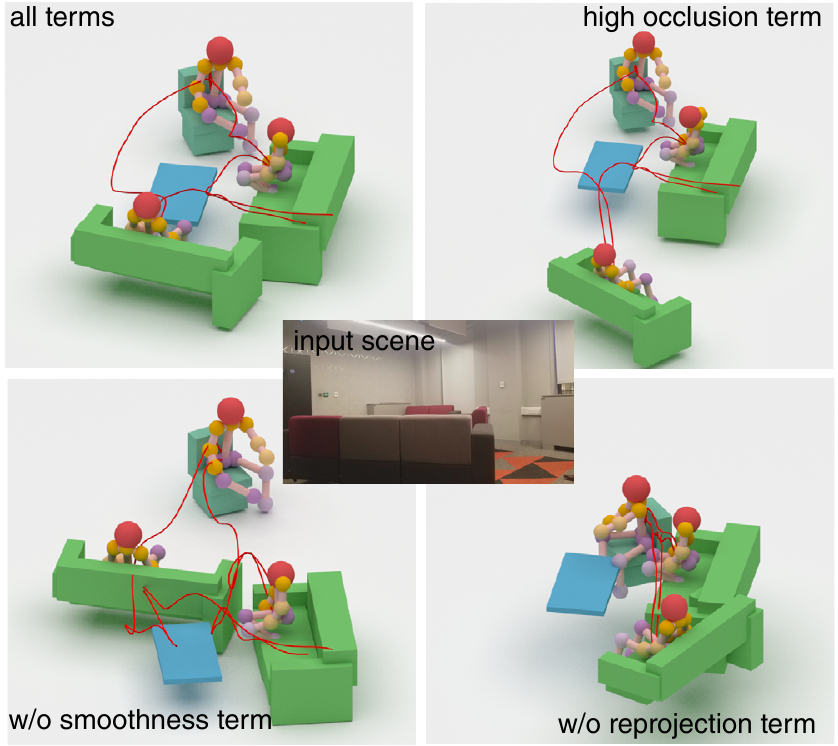}
  	\Large
  	\put(41.5,85){\color{black}\textbf{\Scene{lobby22-1}}}
  \end{overpic}
 % \begin{overpic}[width=\columnwidth]{images/mask-rcnn/lobby22-1-00112.png}
 % 	\Large
  %	\put(2,50){\color{white}\textbf{\maskrcnn} \protect{\cite{He:2017:ICCV}}}
  %\end{overpic}
%   \vspace{2in}
  \caption{We test the effect of various terms on the solution for the scene shown in the center. Reprojection and smoothness generally have the largest influence, since they are used throughout the entire pipeline. }
  \label{fig:ablation}
  \vspace{-10pt}
\end{figure}

In Figure~\ref{fig:ablation} we report the effect of removing some of the terms from our optimization. The reprojection and smoothness terms have the highest effect on the result, since they are used throughout the entire pipeline. Omitting the smoothness term allows for strong path deformations, while omitting the reprojection term removes the anchoring of the motions to the video and permits the smoothness term to strongly contract the path.
The other terms have a more situational effect on the results: how much they influence the result depends on the scene configuration. The occlusion term, for example, has a heavy influence if multiple joint locations are occluded in the video.
Recall that our occlusion term is assymmetric, so that more occlusion always has less cost. The couch is thus optimized to be closer to the camera to occlude more of the scene (the camera is at the bottom center of the image).

%\mypara{qualitative versus quantitative evaluation.}
%a. 2D keypoints with synthetic occlusion
%2D input (not video) -> 3D pose + object locations .. evaluate consistency + plausibility;
%b. test on synthetic data?

%%
%% Quantitative
%%

%qualitative vs quantitative

%comment about im2cad or marr revisited

%ablation

\revision{
\mypara{Hold-one-out validation} We evaluated our method on a hold out-set taken from the PiGraphs dataset. We pick a single scene and remove the scenelets that were generated from this scene from our database, accounting for $\sim 10\%$ of our scenelets.
%a single scene in the PiGraphs dataset. In this experiment, 
%by removing this scene from our scenelet database. 
%As a second experiment on an out-of-house scene, we picked a scene from the PiGraph dataset, and left out the \~10\% scenelets that were generated from this scene from the database.
We compare to the groundtruth objects given in the PiGraphs dataset through manually established correspondence.
%to estimate the top-view distance of object centroids (as described in \Cref{ss:comparisons}).
Our mean reconstruction error was 110 cm (std: 56 cm) with all relevant objects detected.
}

\subsection{Limitations}
We currently assume the input camera to be fixed during the entire duration of the capture. Such a fixed viewpoint results in heavy occlusion in crowded scenes and makes the job of \name\ unnecessarily challenging. In large-scale environments, one can imagine having a network of fixed cameras to address this issue and jointly process the information from the individual cameras.
An obvious limitation of \name\ is failing to react when it `sees' an interaction that is missing in its interaction database. This is an unavoidable problem in any data-driven approach unless we are able to additionally synthesize new interactions, which is a significantly more challenging problem.
%
%Furthermore, since we only rely on detected 2D joint locations from the monocular video, our method will fail when the 2D location detections are erroneous or if too much data is missing.
% partial signals can be confused with occlusions
%have only partial signals.
%
Finally, since our method builds on the expectation that people react similarly in similar settings, it will naturally get confused when this assumption is broken. For example, if a person decides to hand-walk, or use a sofa as a bed, \textit{etc}. We expect this to be partially addressed by richer interaction databases with associated probability priors on actions, and spatio-temporal reasoning that detects and ignores `outlier' interactions by observing a scene over longer time intervals.

%% file: 060_table_performance.inc
\begin{table}[h!]
    \centering
    \caption{Statistics of scenes presented in the paper showing frame count, fraction of frames with occlusion (frac. frames), number of objects in the scene with interactions (objs.). For comparison, we list number of objects detected by Mask-RCNN (MR) while \name detected {\em all} the objects with interactions. Also, we show quality of depth estimation error in cm by MonoDepth (MonoD) and \name as mean (s.d.) $(\mu (\sigma))$ compared against groundtruth annotations.}
    \small
    \begin{tabular}{l|r|r|r||r||r|r}
        scene & frame & frac. & objs. & MR & MonoD & \name \\
         & \# & frames &       &        & $\mu (\sigma)$ & $\mu (\sigma)$\\
        \hline
        \hline
        \Scene{office1-1}    \if0\Cref{fig:teaser}                    \fi & 348 & 1.00 & 7 & 5  & 263 (117) & \textbf{81} (43) \\
        \Scene{livingroom00} \if0\Cref{fig:top_view}                  \fi & 380 & 0.84 & 5 & 4  & 98 (36) & \textbf{57} (32)\\
        \Scene{library3}            									  & 539 & 0.49 & 6 & 5  & 174 (73) & \textbf{83} (29)\\
        \Scene{garden1}             									  & 430 & 0.86 & 5& 5   & 242 (68) & \textbf{58} (28)\\
        \Scene{lobby15}            									      & 254 & 0.65 & 3 & 1  & 243 (98) & \textbf{70} (55)\\ % lobby15
        \Scene{lobby22-1}          										  & 224 & 0.97 & 4 & 1  & 259 (101) & \textbf{51} (21)\\ % lobby22-1
        \Scene{lobby11}      \if0\Cref{fig:qualitativeObject}, row \#1\fi &  80  & 0.00 & 1 & 2 & 120 (-) & \textbf{45} (-)\\
        \Scene{lobby12} 	 \if0\Cref{fig:qualitativeObject}, row \#2\fi & 130  & 0.57 & 2 & 5   & 120 (10) & \textbf{72} (6)\\
        \Scene{lobby18-1}    \if0\Cref{fig:qualitativeObject}, row \#3\fi & 120  & 0.82 & 1 & 1  & 191 (-) & \textbf{25} (-)\\
        \Scene{lobby19-3} 	 \if0\Cref{fig:qualitativeObject}, row \#4\fi & 115  & 0.77 & 2 & 2  & 203 (12) & \textbf{70} (57)\\
        \Scene{lobby24-3-3}  \if0\Cref{fig:scene_evolution}, topLt.   \fi &  49  & 0.84 & 2 & 2  & 197 (24) & \textbf{135} (13)\\
        \Scene{lobby24-3-1}  \if0\Cref{fig:scene_evolution}, topRt.   \fi & 148  & 0.93 & 3 & 2  & 230 (52) & \textbf{72} (41)\\
        \Scene{lobby24-3-2}  \if0\Cref{fig:scene_evolution}, botLt.   \fi & 182  & 1.00 & 4 & 2  & 229 (48) & \textbf{53} (49)\\
        \Scene{lobby24-2-2}  \if0\Cref{fig:scene_evolution}, botRt.   \fi & 189  & 0.98 & 4 & 2  & 216 (50) & \textbf{69} (38)
    \end{tabular}
    \label{tab:occlusions}
\end{table}

%% file: 062_quant_joints.tex
\setlength{\tabcolsep}{0.11em}
\begin{table}[t!]
	\small
	\caption{ Comparison of pose estimates by \name against Tome et al.~\shortcite{tome2017lifting}, \lcrnetthreed~\cite{RogezWS18}, and VNect~\cite{VNect_SIGGRAPH2017}. Note that Tome et al. and \lcrnetthreed return only image-space and local coordinates~(LC); while, VNect is the only other method returning world coordinates~(WC). All units are in cm. 
	}
	\centering
	\begin{tabular}{l||ccc||ccc||ccc||ccc}
		\hline 
		&	\multicolumn{3}{|c||}{ {\tomethreed}}	&	\multicolumn{3}{|c||}{ {\lcrnetthreed}}	& \multicolumn{3}{|c||}{ {Vnect}} &	\multicolumn{3}{|c}{ {\name}}	\\
		\hline 
		&	\multicolumn{1}{|c|}{ {WC}}	&	\multicolumn{1}{|c|}{ {LC}}	&	\multicolumn{1}{|c||}{ {2D}}	  &	\multicolumn{1}{|c|}{ {WC}}	&	\multicolumn{1}{|c|}{ {LC}}	&	\multicolumn{1}{|c||}{ {2D}}	  &	\multicolumn{1}{|c|}{ {WC}}	&	\multicolumn{1}{|c|}{ {LC}}	&	\multicolumn{1}{|c||}{ {2D}}	& 
		\multicolumn{1}{|c|}{ {WC}}	&	\multicolumn{1}{|c|}{ {LC}}	&	\multicolumn{1}{|c }{ {2D}}	\\
		\hline \hline
     {\Scene{livingroom00}}	&	 x	&	 {22.1}	&	 {346.0}	&	 x	&	 {\textbf{12.0}}	&	 {\textbf{77.0}}	& x & x & x &	 \textbf{71.5}	&	 {20.9}	&	 {197.6}	\\
	 {\Scene{lobby15}}	&	 x	&	 {28.1}	&	 {122.5}	&	 x	&	 {\textbf{21.9}}	&	 {\textbf{63.6}}	&	x & x & x & {\textbf{60.8}}	&	 {29.2}	&	 {148.9}	\\
	 {\Scene{lobby22-1}}	&	 x	&	 {19.6}	&	 {81.0}	&	 x	&	 {20.1}	&	 {73.6}	&	 	  x & x & x  & {\textbf{32.0}}	& {\textbf{19.2}}	&	 {\textbf{68.6}} 	\\
		 \Scene{lobby18-1}	&	x	&	{22.4}	&	{70.6}
		                    &   x	&	\textbf{19.2}	&	{83.5}
		                    &  x & x & x
		                    & \textbf{56.7}	&	{19.7}	&	{\textbf{65.4}}	\\
	    \Scene{lobby19-3}	&	x	&	{25.1}	&	{91.6}	&	x	&	{20.7}	&	{104.1}	&  x & x & x &	{\textbf{61.8}}	&	{\textbf{13.6}}	&	{\textbf{46.3}}	\\		 
	    \Scene{vnect}	&	 x	&	 35.5	&	237.9	&	 x	&	 \textbf{20.0}	&	 150.4	&	 68.3 & 28.9 &  171.1 	&	 \textbf{40.3} & 23.5 &	\textbf{138.5}	\\
	\hline
	\end{tabular}
	\label{tab:poses}
\end{table}

%% file: 07_discussion.tex
\section{Conclusion and Future Work} \label{sec:discussion}

Based on the observation that humans (both the same person or different people) interact similarly in similar scenes, we presented
\name\ to {\em jointly} reason about
%(static)
static
object arrangements and human movements. 
At the heart of \name\ lies a novel data-driven method to define {\em characteristic} poses that help identify space-time moments when matched human poses provide reliable cues about the surrounding scene arrangement. Our method links such partial scenelets, fitted to monocular video  under unknown occlusion, and assembles them
to form a global scene layout and 3D human pose estimates.
%space-time estimation. 
%
By extensive evaluation we demonstrate that \name\ improves both the quality of scene layout estimation as well as 3D pose estimates, especially in scenes with low to medium levels of occlusions.

Exciting research directions lay ahead as we are only starting to capture, analyze, and understand the space of (human) interactions, or {\em interaction landscapes} (\cf \cite{Pirk:2017}). Below we discuss some of the immediate issues. 

{\noindent \em Capturing richer interaction databases:} Current datasets only capture limited variety of interactions. Limited refers to both different types of interactions, and variance for each interaction type. For example, we miss examples of interactions with small objects (\eg picking up a cup/glass, using pots and pans in kitchens,  lifting a bag or suitcase), or examples of the different ways that people sit in
%many different people sitting on
sofas, couches, chairs, etc. While significant progress has been made in capturing static environments at high geometric detail, capturing interactions  remains {\em fundamentally} difficult due to heavy occlusion arising due to the interactions. One possibility is to separate the capture of static geometry (\eg with mobile 3D scanners) from the capture of interactions using a  
%ly capture scenes  with mobile 3D scanners, and then interactions using a
mix of sensors like IMU sensors, RGBD scanners, markers, \textit{etc}. We expect such data gathering efforts to happen in the near future. 

{\noindent \em Utilizing scene priors:} We used signals only from interactions. However, in scenes with heavy occlusion, scenelet matching with partial (occluded) information is not sufficient to accurately ground object positions. Also, we cannot directly distinguish between objects that afford similar interactions, such as couches and chairs.
%couch versus chair based on only 2D human features.
One direction would be to additionally use scene statistics and local context, as has been heavily utilized in scene synthesis research, to regularize the  interaction-based reconstruction problem.
%scene reconstruction problem from interactions.
%Also, it is possible to perform space-time smoothing of the recovered motion trajectories using extended Kalman filter based approaches. 

{\noindent \em Recovering interactions over large timescales:} As shown in Figure~\ref{fig:scene_evolution}, \name\ has only a chance of recovering scene arrangements once people interact with various parts of the environment. This suggests that the approach gets better as we `observe' the scene over larger timescales, ideally days or weeks. However, then our static scene assumption breaks down as objects are going to be shifted and moved around. Hence, we would like to extend our approach to also capture space-time object movements, starting with rigid movement of objects, such as a moving chair.
%e.g., moving a chair, etc.

\if0
\subsection{Future Work}

\ersin{Please take a look at the paragraph below that I have written based on our skype discussion. This might move into the discussion section rather than future work.}

One can think that a piece-wise linear extended Kalman filter might be used as a post-process after our final results to further ensure a smooth and continuous dynamic trajectory. However, note that our trajectories are optimized to also respect the constraints of the obstacles in the scene. Such constraints cannot be explicitly satisfied by a Kalman filter on the trajectories only. However, one potential improvement using a Kalman filter is on our initial trajectories before the final joint optimization. This might help initialize the problem better. 
\fi

\if0
\section*{Acknowledgements}
Tuanfeng (mocap), Sebastian Friston (Mocap), Veronika (Mocap)
Daniyar (bilinear ipol), Clement (TF), Moos (SUNCG), Michael (tf), Carlo (deconv), Robin (einsum), 
Reading: Dan, Tom
Code: Denis
\fi

%% file: appendix.tex
\section{Multi-person Tracker}
\label{sec:multi_person_anno}

\revision{In case of multi-person videos, we require the user to inspect the detected 2D skeletons over time and mark the ones that need tracking in the first frame they appear. The input pose detections are not temporally consistent, so we perform a naive actor-reidentification in the form of an MRF optimization. Specifically, given the number of actors $n_a$, the set of skeletons $S(t)$ detected in frame $t$ of the video, our goal is to find the values of the binary variables $\xiactor \in \{0, 1\}$ where $\xiactor = 1$ if actor a is associated with skeleton $s \in S(t)$. To allow some actors to be identified as \emph{invisible} in certain portions of the video, we define a dummy skeleton $\bar{s}$ such that if $\xi_{a \rightarrow \bar{s}}^{t} = 1$, $a$ is marked as invisible. We represent the user annotations for a new actor in the first frame they appear as hard constraints in our optimization. In addition, we define unary and binary terms. The unary term, $E_u(a,s,t)$, measures the cost of assigning an actor $a$ to a skeleton $s$ at a frame $t$ and is based on per-joint confidence measures, $c_{k, s}^t$:
\begin{align}
E_u(a,s,t) = - \xiactor 
          \begin{cases} 
            \frac{1}{n_{joints}} \sum_{k}^{n_{joints}}{c_{k, s}^t} & \textit{if } s \neq \bar{s} \\
            1                                                      & \textit{else}.
          \end{cases}
\end{align}

Assigning the dummy skeleton, $\bar{s}$, to any actor results in a fixed cost of $1$.
The binary term, $E_b(a,s_0,s_1,t)$, measures the cost of assigning an actor $a$ to the skeleton $s_0$ and $s_1$ in frames $t$ and $t+1$ respectively:
\begin{align}
E_b(a,s_0,s_1, t) &=  \xiactort{a}{s_0}{t} \xiactort{a}{s_1}{t+1} C(a, s_0, s_1, t) \\
C(a, s_0, s_1, t) &= \nonumber \\
        & \begin{cases}
          1 & \textit{if } \bar{s} \in \{s_0, s_1\} \\
          \frac{1}{n_{joints}}
          \sum_{k}^{n_{joints}}{
           \left\lVert 
            \frac{u_{k,s_0}^t - u_{k,s_1}^{t+1}}{diag}
           \right\rVert^2_2 c_{k, s_0}^t c_{k, s_1}^{t+1}
          } & \textit{else,} \nonumber
        \end{cases}
\end{align}
where $diag$ refers the the half of the diagonal of the image and transitioning from/to the dummy skeleton results in a fixed cost of $1$. Finally, we optimize for the following energy function:

\begin{align}
    \underset{\xi}{\argmin}~{E} =
        % unary (sum of joint confidences, we want the more confident detection out of two at the same place):
        & -\sum_t \sum_{a \leq n_a, s \in S(t)} E_u(a,s,t)
        \nonumber \\
        % pairwise (2D per joint distance, we don't want large screen space jumps):
        & + w_{pw}
        \sum_t \sum_{a \leq n_a}
        \sum_{s_0 \in S(t)}
        \sum_{s_1 \in S(t+1)}
         E_b(a,s_0,s_1, t)
    \label{eq:mrf}
\end{align}
subject to the constraints:
\begin{align}
        %
        % constraints:
        %\bigforall_{a} \bigforall_{s} \bigforall_{t} ~~ \xiactor \in \{0, 1\}\ , \\
        % each actor 'a' in each frame 't' has only one skeleton 's' assigned,
        % ' = 1', because we need an assignment for each actor:
        \bigforall_{t} \bigforall_{s} ~~ \sum_{a \leq n_a}{\xiactor} = 1 \ ,
        % each skeleton 's' in each frame 't' only used for at most one actor 'a',
        % ' <= 1' because not all skeletons are assigned:
        ~~~\bigforall_{t} \bigforall_{a} ~~ \sum_{s \neq \bar{s}}{\xiactor} \leq 1.
        % the number of skeleton detections in each frame varies, and we always add a 
        %& S(t) = \left\{ \left\{u_{k,s}^t\right\}, ~\underbrace{\bar{s}}_{\textit{'not detected'}} \right\}
        %\nonumber 
    \label{eq:mrfConstraints}
\end{align}
We set the relative weight of the binary term in Equation~\ref{eq:mrf} as $w_{pw} = 10^3$ and optimize using a binary discrete optimizer (Gurobi). 

For scenes with lots of occlusion and crossings, we require additional manual constraints. Once, multi-actor/2D skeleton associations are established, we optimize for 3D poses by enforcing smoothness between poses that belong to the same actor only. 

}%revision end

%\begin{itemize}
%    \item Input: 
%      i) number of actors $n_a$, 
%      ii) 2D skeletons (grouped), 
%      iii) [manual hard constraints (actor-skeleton assignments)]
%    \item Output: Smooth skeleton trajectories in time, roughly corresponding to identities.
%    \item Setup: 
%    \item Node: Actor-Skeleton assignment, or Actor is invisible
%    \item ~Unary cost: sum of per-joint confidence (negative for minimization), fixed penalty for invisible
%    \item ~Pairwise cost: sum of 2D joint distances, fixed penalty for transitioning from/to invisible
%\end{itemize}
%} % highlight end
\if0
\hl{Implementation in the pipeline: concatenated poses with removed smoothness terms between actors, and candidate scenelet times (we don't consider fitting a scenelet to half of one actor and half of the other).
\fi
%} % highlight end

\section{Confidence of Keypoint Detection using \lcrnet}
\label{appendix:confidence}
When using \lcrnet, we confidence estimate 2D detection keypoints as
\begin{align}\revision{
    v_k} & \revision{= \frac{\underset{q_k^i \in pose\ proposals}{var}\left(q_k^i\right)}{1 + \exp\left(-0.2 s' + 3.5\right)}} \\
    \revision{c_k (v_k)} &\revision{ = \frac{1}{
      1 + \exp\left(-10 \exp\left(\frac{log(v_k)}{P_{99}\left(log(v_k)\right)}\right) + 24\right),
    }
}\end{align}
\revision{where $var_k$ denotes the variance of the 3D joint position among the grouped pose proposals, and $P_{99}$ denotes the $99$th percentile of $log$ joint variances over the whole recording, assigning high confidence to low variance joint estimates, and $s'$ is a per-pose score defined in Equation 6 in \cite{RogezWS18}.}

%% file: main_arxiv.bbl

\begin{thebibliography}{00}

%%% the Web address.

\ifx \showCODEN    \undefined \def \showCODEN     #1{\unskip}     \fi
\ifx \showDOI      \undefined \def \showDOI       #1{#1}\fi
\ifx \showISBNx    \undefined \def \showISBNx     #1{\unskip}     \fi
\ifx \showISBNxiii \undefined \def \showISBNxiii  #1{\unskip}     \fi
\ifx \showISSN     \undefined \def \showISSN      #1{\unskip}     \fi
\ifx \showLCCN     \undefined \def \showLCCN      #1{\unskip}     \fi
\ifx \shownote     \undefined \def \shownote      #1{#1}          \fi
\ifx \showarticletitle \undefined \def \showarticletitle #1{#1}   \fi
\ifx \showURL      \undefined \def \showURL       {\relax}        \fi
% invisible to TeX
\providecommand\bibfield[2]{#2}
\providecommand\bibinfo[2]{#2}
\providecommand\natexlab[1]{#1}
\providecommand\showeprint[2][]{arXiv:#2}

\bibitem[\protect\citeauthoryear{Brostow and Essa}{Brostow and Essa}{1999}]%
        {brostow1999motion}
\bibfield{author}{\bibinfo{person}{Gabriel~J Brostow} {and}
  \bibinfo{person}{Irfan~A Essa}.} \bibinfo{year}{1999}\natexlab{}.
\newblock \showarticletitle{Motion based decompositing of video}.
\newblock \bibinfo{journal}{{\em {IEEE ICCV}\/}} (\bibinfo{year}{1999}).
\newblock


\bibitem[\protect\citeauthoryear{Cao, Simon, Wei, and Sheikh}{Cao
  et~al\mbox{.}}{2017}]%
        {cao2017}
\bibfield{author}{\bibinfo{person}{Zhe Cao}, \bibinfo{person}{Tomas Simon},
  \bibinfo{person}{Shih-En Wei}, {and} \bibinfo{person}{Yaser Sheikh}.}
  \bibinfo{year}{2017}\natexlab{}.
\newblock \showarticletitle{Realtime Multi-Person 2D Pose Estimation using Part
  Affinity Fields}. In \bibinfo{booktitle}{{\em {IEEE CVPR}}}.
\newblock


\bibitem[\protect\citeauthoryear{Chakrabarti, Shao, and
  Shakhnarovich}{Chakrabarti et~al\mbox{.}}{2016}]%
        {ChakrabartiSS16}
\bibfield{author}{\bibinfo{person}{Ayan Chakrabarti}, \bibinfo{person}{Jingyu
  Shao}, {and} \bibinfo{person}{Gregory Shakhnarovich}.}
  \bibinfo{year}{2016}\natexlab{}.
\newblock \showarticletitle{Depth from a Single Image by Harmonizing
  Overcomplete Local Network Predictions}.
\newblock \bibinfo{journal}{{\em CoRR\/}}  \bibinfo{volume}{abs/1605.07081}
  (\bibinfo{year}{2016}).
\newblock


\bibitem[\protect\citeauthoryear{Chang, Dai, Funkhouser, Halber, Niessner,
  Savva, Song, Zeng, and Zhang}{Chang et~al\mbox{.}}{2017}]%
        {Matterport3D}
\bibfield{author}{\bibinfo{person}{Angel Chang}, \bibinfo{person}{Angela Dai},
  \bibinfo{person}{Thomas Funkhouser}, \bibinfo{person}{Maciej Halber},
  \bibinfo{person}{Matthias Niessner}, \bibinfo{person}{Manolis Savva},
  \bibinfo{person}{Shuran Song}, \bibinfo{person}{Andy Zeng}, {and}
  \bibinfo{person}{Yinda Zhang}.} \bibinfo{year}{2017}\natexlab{}.
\newblock \showarticletitle{Matterport3D: Learning from RGB-D Data in Indoor
  Environments}.
\newblock \bibinfo{journal}{{\em 3DV\/}}.
\newblock


\bibitem[\protect\citeauthoryear{Chen, Lai, Wu, Martin, and Hu}{Chen
  et~al\mbox{.}}{2014}]%
        {Chen:2014}
\bibfield{author}{\bibinfo{person}{Kang Chen}, \bibinfo{person}{Yu-Kun Lai},
  \bibinfo{person}{Yu-Xin Wu}, \bibinfo{person}{Ralph Martin}, {and}
  \bibinfo{person}{Shi-Min Hu}.} \bibinfo{year}{2014}\natexlab{}.
\newblock \showarticletitle{Automatic Semantic Modeling of Indoor Scenes from
  Low-quality RGB-D Data Using Contextual Information}.
\newblock \bibinfo{journal}{{\em {ACM SIGGRAPH Asia}\/}} \bibinfo{volume}{33},
  \bibinfo{number}{6}, Article \bibinfo{articleno}{208} (\bibinfo{date}{Nov.}
  \bibinfo{year}{2014}), \bibinfo{numpages}{12}~pages.
\newblock
\showISSN{0730-0301}


\bibitem[\protect\citeauthoryear{Dai, Chang, Savva, Halber, Funkhouser, and
  Nie{\ss}ner}{Dai et~al\mbox{.}}{2017a}]%
        {dai2017scannet}
\bibfield{author}{\bibinfo{person}{Angela Dai}, \bibinfo{person}{Angel~X.
  Chang}, \bibinfo{person}{Manolis Savva}, \bibinfo{person}{Maciej Halber},
  \bibinfo{person}{Thomas Funkhouser}, {and} \bibinfo{person}{Matthias
  Nie{\ss}ner}.} \bibinfo{year}{2017}\natexlab{a}.
\newblock \showarticletitle{ScanNet: Richly-annotated 3D Reconstructions of
  Indoor Scenes}. In \bibinfo{booktitle}{{\em {IEEE CVPR}}}.
\newblock


\bibitem[\protect\citeauthoryear{Dai, Nie{\ss}ner, Zoll{\"o}fer, Izadi, and
  Theobalt}{Dai et~al\mbox{.}}{2017b}]%
        {dai2017bundlefusion}
\bibfield{author}{\bibinfo{person}{Angela Dai}, \bibinfo{person}{Matthias
  Nie{\ss}ner}, \bibinfo{person}{Michael Zoll{\"o}fer},
  \bibinfo{person}{Shahram Izadi}, {and} \bibinfo{person}{Christian Theobalt}.}
  \bibinfo{year}{2017}\natexlab{b}.
\newblock \showarticletitle{BundleFusion: Real-time Globally Consistent 3D
  Reconstruction using On-the-fly Surface Re-integration}.
\newblock \bibinfo{journal}{{\em {ACM TOG}\/}} (\bibinfo{year}{2017}).
\newblock


\bibitem[\protect\citeauthoryear{Del~Pero, Bowdish, Kermgard, Hartley, and
  Barnard}{Del~Pero et~al\mbox{.}}{2013}]%
        {Pero_2013_CVPR}
\bibfield{author}{\bibinfo{person}{Luca Del~Pero}, \bibinfo{person}{Joshua
  Bowdish}, \bibinfo{person}{Bonnie Kermgard}, \bibinfo{person}{Emily Hartley},
  {and} \bibinfo{person}{Kobus Barnard}.} \bibinfo{year}{2013}\natexlab{}.
\newblock \showarticletitle{Understanding Bayesian Rooms Using Composite 3D
  Object Models}. In \bibinfo{booktitle}{{\em {IEEE CVPR}}}.
\newblock


\bibitem[\protect\citeauthoryear{Delaitre, Fouhey, Laptev, Sivic, Gupta, and
  Efros}{Delaitre et~al\mbox{.}}{2012}]%
        {delaitre2012}
\bibfield{author}{\bibinfo{person}{V. Delaitre}, \bibinfo{person}{D. Fouhey},
  \bibinfo{person}{I. Laptev}, \bibinfo{person}{J. Sivic}, \bibinfo{person}{A.
  Gupta}, {and} \bibinfo{person}{A. Efros}.} \bibinfo{year}{2012}\natexlab{}.
\newblock \showarticletitle{Scene semantics from long-term observation of
  people}.
\newblock \bibinfo{journal}{{\em {ECCV}\/}} (\bibinfo{year}{2012}).
\newblock


\bibitem[\protect\citeauthoryear{Fisher, Ritchie, Savva, Funkhouser, and
  Hanrahan}{Fisher et~al\mbox{.}}{2012}]%
        {2012-scenesynth}
\bibfield{author}{\bibinfo{person}{Matthew Fisher}, \bibinfo{person}{Daniel
  Ritchie}, \bibinfo{person}{Manolis Savva}, \bibinfo{person}{Thomas
  Funkhouser}, {and} \bibinfo{person}{Pat Hanrahan}.}
  \bibinfo{year}{2012}\natexlab{}.
\newblock \showarticletitle{Example-based Synthesis of 3D Object Arrangements}.
  In \bibinfo{booktitle}{{\em {ACM SIGGRAPH Asia}}}.
\newblock


\bibitem[\protect\citeauthoryear{Fisher, Savva, and Hanrahan}{Fisher
  et~al\mbox{.}}{2011}]%
        {fisher2011}
\bibfield{author}{\bibinfo{person}{Matthew Fisher}, \bibinfo{person}{Manolis
  Savva}, {and} \bibinfo{person}{Pat Hanrahan}.}
  \bibinfo{year}{2011}\natexlab{}.
\newblock \showarticletitle{Characterizing structural relationships in scenes
  using graph kernels}. In \bibinfo{booktitle}{{\em {ACM SIGGRAPH}}},
  Vol.~\bibinfo{volume}{30}. \bibinfo{pages}{34}.
\newblock


\bibitem[\protect\citeauthoryear{Fisher, Savva, Li, Hanrahan, and
  Nie{\ss}ner}{Fisher et~al\mbox{.}}{2015}]%
        {fisher2015actsynth}
\bibfield{author}{\bibinfo{person}{Matthew Fisher}, \bibinfo{person}{Manolis
  Savva}, \bibinfo{person}{Yangyan Li}, \bibinfo{person}{Pat Hanrahan}, {and}
  \bibinfo{person}{Matthias Nie{\ss}ner}.} \bibinfo{year}{2015}\natexlab{}.
\newblock \showarticletitle{Activity-centric Scene Synthesis for Functional 3D
  Scene Modeling}.
\newblock \bibinfo{journal}{{\em {ACM SIGGRAPH Asia}\/}} \bibinfo{volume}{34},
  \bibinfo{number}{6} (\bibinfo{year}{2015}).
\newblock


\bibitem[\protect\citeauthoryear{Fouhey, Delaitre, Gupta, Efros, Laptev, and
  Sivic}{Fouhey et~al\mbox{.}}{2012}]%
        {Fouhey:PeopleWatching:12}
\bibfield{author}{\bibinfo{person}{David~F. Fouhey}, \bibinfo{person}{Vincent
  Delaitre}, \bibinfo{person}{Abhinav Gupta}, \bibinfo{person}{Alexei~A.
  Efros}, \bibinfo{person}{Ivan Laptev}, {and} \bibinfo{person}{Josef Sivic}.}
  \bibinfo{year}{2012}\natexlab{}.
\newblock \showarticletitle{People Watching: Human Actions as a Cue for
  Single-View Geometry}.
\newblock \bibinfo{journal}{{\em {ECCV}\/}} (\bibinfo{year}{2012}).
\newblock


\bibitem[\protect\citeauthoryear{Frank, Ruhnke, Tatarchenko, and Burgard}{Frank
  et~al\mbox{.}}{2015}]%
        {Frank2015}
\bibfield{author}{\bibinfo{person}{B. Frank}, \bibinfo{person}{M. Ruhnke},
  \bibinfo{person}{M. Tatarchenko}, {and} \bibinfo{person}{W. Burgard}.}
  \bibinfo{year}{2015}\natexlab{}.
\newblock \showarticletitle{3D-reconstruction of indoor environments from human
  activity}. In \bibinfo{booktitle}{{\em IEEE ICRA}}.
  \bibinfo{pages}{4644--4649}.
\newblock


\bibitem[\protect\citeauthoryear{Fu, Zhang, and Huang}{Fu
  et~al\mbox{.}}{2015}]%
        {Fu_2015_ICCV}
\bibfield{author}{\bibinfo{person}{Lianrui Fu}, \bibinfo{person}{Junge Zhang},
  {and} \bibinfo{person}{Kaiqi Huang}.} \bibinfo{year}{2015}\natexlab{}.
\newblock \showarticletitle{Beyond Tree Structure Models: A New Occlusion Aware
  Graphical Model for Human Pose Estimation}. In \bibinfo{booktitle}{{\em {IEEE
  ICCV}}}.
\newblock


\bibitem[\protect\citeauthoryear{Fu, Chen, Su, and Fu}{Fu
  et~al\mbox{.}}{2017a}]%
        {Fu:2017tvcg}
\bibfield{author}{\bibinfo{person}{Q. Fu}, \bibinfo{person}{X. Chen},
  \bibinfo{person}{X. Su}, {and} \bibinfo{person}{H. Fu}.}
  \bibinfo{year}{2017}\natexlab{a}.
\newblock \showarticletitle{Pose-Inspired Shape Synthesis and Functional
  Hybrid}.
\newblock \bibinfo{journal}{{\em {IEEE TVCG}\/}} \bibinfo{volume}{23},
  \bibinfo{number}{12} (\bibinfo{year}{2017}), \bibinfo{pages}{2574--2585}.
\newblock
\showISSN{1077-2626}


\bibitem[\protect\citeauthoryear{Fu, Chen, Wang, Wen, Zhou, and Fu}{Fu
  et~al\mbox{.}}{2017b}]%
        {fu_siga17}
\bibfield{author}{\bibinfo{person}{Qiang Fu}, \bibinfo{person}{Xiaowu Chen},
  \bibinfo{person}{Xiaotian Wang}, \bibinfo{person}{Sijia Wen},
  \bibinfo{person}{Bin Zhou}, {and} \bibinfo{person}{Hongbo Fu}.}
  \bibinfo{year}{2017}\natexlab{b}.
\newblock \showarticletitle{Adaptive Synthesis of Indoor Scenes via
  Activity-Associated Object Relation Graphs}.
\newblock \bibinfo{journal}{{\em {ACM SIGGRAPH Asia}\/}} \bibinfo{volume}{36},
  \bibinfo{number}{6} (\bibinfo{year}{2017}), \bibinfo{pages}{Article No. 201}.
\newblock


\bibitem[\protect\citeauthoryear{Gkioxari, Girshick, Doll\'{a}r, and
  He}{Gkioxari et~al\mbox{.}}{2018}]%
        {gkioxari2017}
\bibfield{author}{\bibinfo{person}{Georgia Gkioxari}, \bibinfo{person}{Ross
  Girshick}, \bibinfo{person}{Piotr Doll\'{a}r}, {and} \bibinfo{person}{Kaiming
  He}.} \bibinfo{year}{2018}\natexlab{}.
\newblock \showarticletitle{Detecting and Recognizing Human-Object
  Intaractions}.
\newblock \bibinfo{journal}{{\em CVPR\/}} (\bibinfo{year}{2018}).
\newblock


\bibitem[\protect\citeauthoryear{Gupta, Kembhavi, and Davis}{Gupta
  et~al\mbox{.}}{2009}]%
        {Gupta:2009:OHI:1608576.1608766}
\bibfield{author}{\bibinfo{person}{Abhinav Gupta}, \bibinfo{person}{Aniruddha
  Kembhavi}, {and} \bibinfo{person}{Larry~S. Davis}.}
  \bibinfo{year}{2009}\natexlab{}.
\newblock \showarticletitle{Observing Human-Object Interactions: Using Spatial
  and Functional Compatibility for Recognition}.
\newblock \bibinfo{journal}{{\em {IEEE PAMI}\/}} \bibinfo{volume}{31},
  \bibinfo{number}{10} (\bibinfo{date}{Oct.} \bibinfo{year}{2009}),
  \bibinfo{pages}{1775--1789}.
\newblock
\showISSN{0162-8828}


\bibitem[\protect\citeauthoryear{He, Gkioxari, Dollár, and Girshick}{He
  et~al\mbox{.}}{2017}]%
        {He:2017:ICCV}
\bibfield{author}{\bibinfo{person}{K. He}, \bibinfo{person}{G. Gkioxari},
  \bibinfo{person}{P. Dollár}, {and} \bibinfo{person}{R. Girshick}.}
  \bibinfo{year}{2017}\natexlab{}.
\newblock \showarticletitle{Mask R-CNN}. In \bibinfo{booktitle}{{\em 2017 IEEE
  International Conference on Computer Vision (ICCV)}}.
  \bibinfo{pages}{2980--2988}.
\newblock
\showDOI{%
\url{https://doi.org/10.1109/ICCV.2017.322}}


\bibitem[\protect\citeauthoryear{Huang, Boyer, Navab, and Ilic}{Huang
  et~al\mbox{.}}{2014}]%
        {HuangKeyframes2014CVPR}
\bibfield{author}{\bibinfo{person}{C.~H. Huang}, \bibinfo{person}{E. Boyer},
  \bibinfo{person}{N. Navab}, {and} \bibinfo{person}{S. Ilic}.}
  \bibinfo{year}{2014}\natexlab{}.
\newblock \showarticletitle{Human Shape and Pose Tracking Using Keyframes}. In
  \bibinfo{booktitle}{{\em {IEEE CVPR}}}. \bibinfo{pages}{3446--3453}.
\newblock
\showISSN{1063-6919}
\showDOI{%
\url{https://doi.org/10.1109/CVPR.2014.440}}


\bibitem[\protect\citeauthoryear{Huang and Yangc}{Huang and Yangc}{2009}]%
        {Huang:2009}
\bibfield{author}{\bibinfo{person}{Jia-Bin Huang} {and}
  \bibinfo{person}{Ming-Hsuan Yangc}.} \bibinfo{year}{2009}\natexlab{}.
\newblock \showarticletitle{Estimating Human Pose from Occluded Images}. In
  \bibinfo{booktitle}{{\em ACCV}}.
\newblock


\bibitem[\protect\citeauthoryear{Huang, Fu, and Hu}{Huang
  et~al\mbox{.}}{2016}]%
        {HUANG201646}
\bibfield{author}{\bibinfo{person}{Shi-Sheng Huang}, \bibinfo{person}{Hongbo
  Fu}, {and} \bibinfo{person}{Shi-Min Hu}.} \bibinfo{year}{2016}\natexlab{}.
\newblock \showarticletitle{Structure guided interior scene synthesis via graph
  matching}.
\newblock \bibinfo{journal}{{\em Graphical Models\/}}  \bibinfo{volume}{85}
  (\bibinfo{year}{2016}), \bibinfo{pages}{46 -- 55}.
\newblock
\showISSN{1524-0703}


\bibitem[\protect\citeauthoryear{Insafutdinov, Pishchulin, Andres, Andriluka,
  and Schiele}{Insafutdinov et~al\mbox{.}}{2016}]%
        {Insafutdinov:DeeperCut:2016}
\bibfield{author}{\bibinfo{person}{Eldar Insafutdinov}, \bibinfo{person}{Leonid
  Pishchulin}, \bibinfo{person}{Bjoern Andres}, \bibinfo{person}{Mykhaylo
  Andriluka}, {and} \bibinfo{person}{Bernt Schiele}.}
  \bibinfo{year}{2016}\natexlab{}.
\newblock \showarticletitle{{DeeperCut: A Deeper, Stronger, and Faster
  Multi-Person Pose Estimation Model}}.
\newblock \bibinfo{journal}{{\em {ECCV}\/}} (\bibinfo{date}{Oct}
  \bibinfo{year}{2016}).
\newblock


\bibitem[\protect\citeauthoryear{Izadinia, Shan, and Seitz}{Izadinia
  et~al\mbox{.}}{2017}]%
        {izadinia2017im2cad}
\bibfield{author}{\bibinfo{person}{Hamid Izadinia}, \bibinfo{person}{Qi Shan},
  {and} \bibinfo{person}{Steven~M Seitz}.} \bibinfo{year}{2017}\natexlab{}.
\newblock \showarticletitle{IM2CAD}. In \bibinfo{booktitle}{{\em CVPR}}.
\newblock


\bibitem[\protect\citeauthoryear{Jiang, Koppula, and Saxena}{Jiang
  et~al\mbox{.}}{2016}]%
        {Jiang:2016}
\bibfield{author}{\bibinfo{person}{Yun Jiang}, \bibinfo{person}{Hema~S.
  Koppula}, {and} \bibinfo{person}{Ashutosh Saxena}.}
  \bibinfo{year}{2016}\natexlab{}.
\newblock \showarticletitle{Modeling 3D Environments Through Hidden Human
  Context}.
\newblock \bibinfo{journal}{{\em {IEEE PAMI}\/}} \bibinfo{volume}{38},
  \bibinfo{number}{10} (\bibinfo{date}{Oct.} \bibinfo{year}{2016}),
  \bibinfo{pages}{2040--2053}.
\newblock
\showISSN{0162-8828}


\bibitem[\protect\citeauthoryear{Kang and Lee}{Kang and Lee}{2017}]%
        {KANG201725}
\bibfield{author}{\bibinfo{person}{Changgu Kang} {and}
  \bibinfo{person}{Sung-Hee Lee}.} \bibinfo{year}{2017}\natexlab{}.
\newblock \showarticletitle{Scene reconstruction and analysis from motion}.
\newblock \bibinfo{journal}{{\em Graphical Models\/}}  \bibinfo{volume}{94}
  (\bibinfo{year}{2017}), \bibinfo{pages}{25 -- 37}.
\newblock
\showISSN{1524-0703}


\bibitem[\protect\citeauthoryear{Kim, Chaudhuri, Guibas, and Funkhouser}{Kim
  et~al\mbox{.}}{2014}]%
        {Kim:2014:SHS}
\bibfield{author}{\bibinfo{person}{Vladimir~G. Kim},
  \bibinfo{person}{Siddhartha Chaudhuri}, \bibinfo{person}{Leonidas Guibas},
  {and} \bibinfo{person}{Thomas Funkhouser}.} \bibinfo{year}{2014}\natexlab{}.
\newblock \showarticletitle{{Shape2Pose}: Human-Centric Shape Analysis}.
\newblock \bibinfo{journal}{{\em {ACM SIGGRAPH}\/}} (\bibinfo{date}{Aug.}
  \bibinfo{year}{2014}).
\newblock


\bibitem[\protect\citeauthoryear{Krasner}{Krasner}{2013}]%
        {principles13}
\bibfield{author}{\bibinfo{person}{Leonard Krasner}.}
  \bibinfo{year}{2013}\natexlab{}.
\newblock \bibinfo{booktitle}{{\em Environmental Design and Human Behavior}}.
\newblock \bibinfo{publisher}{Elsevier}.
\newblock


\bibitem[\protect\citeauthoryear{Liu, Chaudhuri, Kim, Huang, Mitra, and
  Funkhouser}{Liu et~al\mbox{.}}{2014}]%
        {Liu:2014}
\bibfield{author}{\bibinfo{person}{Tianqiang Liu}, \bibinfo{person}{Siddhartha
  Chaudhuri}, \bibinfo{person}{Vladimir~G. Kim}, \bibinfo{person}{Qixing
  Huang}, \bibinfo{person}{Niloy~J. Mitra}, {and} \bibinfo{person}{Thomas
  Funkhouser}.} \bibinfo{year}{2014}\natexlab{}.
\newblock \showarticletitle{Creating Consistent Scene Graphs Using a
  Probabilistic Grammar}.
\newblock \bibinfo{journal}{{\em {ACM SIGGRAPH Asia}\/}} \bibinfo{volume}{33},
  \bibinfo{number}{6}, Article \bibinfo{articleno}{211} (\bibinfo{date}{Nov.}
  \bibinfo{year}{2014}), \bibinfo{numpages}{12}~pages.
\newblock
\showISSN{0730-0301}


\bibitem[\protect\citeauthoryear{Ma, Li, Zou, Liao, Tong, and Zhang}{Ma
  et~al\mbox{.}}{2016}]%
        {Ma:2016:AIS:2980179.2980223}
\bibfield{author}{\bibinfo{person}{Rui Ma}, \bibinfo{person}{Honghua Li},
  \bibinfo{person}{Changqing Zou}, \bibinfo{person}{Zicheng Liao},
  \bibinfo{person}{Xin Tong}, {and} \bibinfo{person}{Hao Zhang}.}
  \bibinfo{year}{2016}\natexlab{}.
\newblock \showarticletitle{Action-driven 3D Indoor Scene Evolution}.
\newblock \bibinfo{journal}{{\em {ACM SIGGRAPH Asia}\/}} \bibinfo{volume}{35},
  \bibinfo{number}{6}, Article \bibinfo{articleno}{173} (\bibinfo{date}{Nov.}
  \bibinfo{year}{2016}), \bibinfo{numpages}{13}~pages.
\newblock
\showISSN{0730-0301}


\bibitem[\protect\citeauthoryear{Mehta, Rhodin, Casas, Fua, Sotnychenko, Xu,
  and Theobalt}{Mehta et~al\mbox{.}}{2017a}]%
        {mono-3dhp2017}
\bibfield{author}{\bibinfo{person}{Dushyant Mehta}, \bibinfo{person}{Helge
  Rhodin}, \bibinfo{person}{Dan Casas}, \bibinfo{person}{Pascal Fua},
  \bibinfo{person}{Oleksandr Sotnychenko}, \bibinfo{person}{Weipeng Xu}, {and}
  \bibinfo{person}{Christian Theobalt}.} \bibinfo{year}{2017}\natexlab{a}.
\newblock \showarticletitle{Monocular 3D Human Pose Estimation In The Wild
  Using Improved CNN Supervision}. In \bibinfo{booktitle}{{\em 3D Vision (3DV),
  2017 Fifth International Conference on}}. IEEE.
\newblock
\showDOI{%
\url{https://doi.org/10.1109/3dv.2017.00064}}


\bibitem[\protect\citeauthoryear{Mehta, Sridhar, Sotnychenko, Rhodin, Shafiei,
  Seidel, Xu, Casas, and Theobalt}{Mehta et~al\mbox{.}}{2017b}]%
        {VNect_SIGGRAPH2017}
\bibfield{author}{\bibinfo{person}{Dushyant Mehta}, \bibinfo{person}{Srinath
  Sridhar}, \bibinfo{person}{Oleksandr Sotnychenko}, \bibinfo{person}{Helge
  Rhodin}, \bibinfo{person}{Mohammad Shafiei}, \bibinfo{person}{Hans-Peter
  Seidel}, \bibinfo{person}{Weipeng Xu}, \bibinfo{person}{Dan Casas}, {and}
  \bibinfo{person}{Christian Theobalt}.} \bibinfo{year}{2017}\natexlab{b}.
\newblock \showarticletitle{VNect: Real-time 3D Human Pose Estimation with a
  Single RGB Camera}.
\newblock \bibinfo{journal}{{\em {ACM SIGGRAPH}\/}} \bibinfo{volume}{36},
  \bibinfo{number}{4}, 14.
\newblock


\bibitem[\protect\citeauthoryear{Nan, Xie, and Sharf}{Nan
  et~al\mbox{.}}{2012}]%
        {Nan:2012}
\bibfield{author}{\bibinfo{person}{Liangliang Nan}, \bibinfo{person}{Ke Xie},
  {and} \bibinfo{person}{Andrei Sharf}.} \bibinfo{year}{2012}\natexlab{}.
\newblock \showarticletitle{A Search-classify Approach for Cluttered Indoor
  Scene Understanding}.
\newblock \bibinfo{journal}{{\em {ACM SIGGRAPH Asia}\/}} \bibinfo{volume}{31},
  \bibinfo{number}{6}, Article \bibinfo{articleno}{137} (\bibinfo{date}{Nov.}
  \bibinfo{year}{2012}), \bibinfo{numpages}{10}~pages.
\newblock
\showISSN{0730-0301}


\bibitem[\protect\citeauthoryear{Neisser}{Neisser}{1976}]%
        {psyc76}
\bibfield{author}{\bibinfo{person}{Ulric Neisser}.}
  \bibinfo{year}{1976}\natexlab{}.
\newblock \bibinfo{booktitle}{{\em Environmental Design and Human Behavior}}.
\newblock \bibinfo{publisher}{W. H. Freeman, SF}.
\newblock


\bibitem[\protect\citeauthoryear{Newcombe et~al\mbox{.}}{Newcombe
  et~al\mbox{.}}{2011}]%
        {Newcombe:2011:KRD}
\bibfield{author}{\bibinfo{person}{Richard~A. Newcombe} {et~al\mbox{.}}}
  \bibinfo{year}{2011}\natexlab{}.
\newblock \showarticletitle{KinectFusion: Real-time dense surface mapping and
  tracking}. In \bibinfo{booktitle}{{\em IEEE ISMAR}}.
  \bibinfo{pages}{127--136}.
\newblock


\bibitem[\protect\citeauthoryear{Newcombe, Fox, and Seitz}{Newcombe
  et~al\mbox{.}}{2015}]%
        {Newcombe2015DynamicFusionRA}
\bibfield{author}{\bibinfo{person}{Richard~A. Newcombe},
  \bibinfo{person}{Dieter Fox}, {and} \bibinfo{person}{Steven~M. Seitz}.}
  \bibinfo{year}{2015}\natexlab{}.
\newblock \showarticletitle{DynamicFusion: Reconstruction and tracking of
  non-rigid scenes in real-time}.
\newblock \bibinfo{journal}{{\em {IEEE CVPR}\/}} (\bibinfo{year}{2015}).
\newblock


\bibitem[\protect\citeauthoryear{Newell, Yang, and Deng}{Newell
  et~al\mbox{.}}{2016}]%
        {Newell:PoseHg:2016}
\bibfield{author}{\bibinfo{person}{Alejandro Newell}, \bibinfo{person}{Kaiyu
  Yang}, {and} \bibinfo{person}{Jia Deng}.} \bibinfo{year}{2016}\natexlab{}.
\newblock \showarticletitle{Stacked Hourglass Networks for Human Pose
  Estimation}.
\newblock \bibinfo{journal}{{\em {ECCV}\/}} (\bibinfo{year}{2016}).
\newblock


\bibitem[\protect\citeauthoryear{Pirk, Diamanti, Thibert, Xu, and Guibas}{Pirk
  et~al\mbox{.}}{2017a}]%
        {Pirk:2017icip}
\bibfield{author}{\bibinfo{person}{S\"{o}ren Pirk}, \bibinfo{person}{Olga
  Diamanti}, \bibinfo{person}{Boris Thibert}, \bibinfo{person}{Danfei Xu},
  {and} \bibinfo{person}{Leonidas~J. Guibas}.}
  \bibinfo{year}{2017}\natexlab{a}.
\newblock \showarticletitle{SHAPE-AWARE SPATIO-TEMPORAL DESCRIPTORS FOR
  INTERACTION CLASSIFICATION}.
\newblock \bibinfo{journal}{{\em ICIP\/}} (\bibinfo{year}{2017}).
\newblock


\bibitem[\protect\citeauthoryear{Pirk, Krs, Hu, Rajasekaran, Kang, Yoshiyasu,
  Benes, and Guibas}{Pirk et~al\mbox{.}}{2017b}]%
        {Pirk:2017}
\bibfield{author}{\bibinfo{person}{S\"{o}ren Pirk}, \bibinfo{person}{Vojtech
  Krs}, \bibinfo{person}{Kaimo Hu}, \bibinfo{person}{Suren~Deepak Rajasekaran},
  \bibinfo{person}{Hao Kang}, \bibinfo{person}{Yusuke Yoshiyasu},
  \bibinfo{person}{Bedrich Benes}, {and} \bibinfo{person}{Leonidas~J. Guibas}.}
  \bibinfo{year}{2017}\natexlab{b}.
\newblock \showarticletitle{Understanding and Exploiting Object Interaction
  Landscapes}.
\newblock \bibinfo{journal}{{\em {ACM SIGGRAPH Asia}\/}} \bibinfo{volume}{36},
  \bibinfo{number}{3}, Article \bibinfo{articleno}{31} (\bibinfo{date}{June}
  \bibinfo{year}{2017}), \bibinfo{numpages}{14}~pages.
\newblock
\showISSN{0730-0301}


\bibitem[\protect\citeauthoryear{Poirson, Ammirato, Fu, Liu, Kosecka, and
  Berg}{Poirson et~al\mbox{.}}{2016}]%
        {Poirson2016}
\bibfield{author}{\bibinfo{person}{P. Poirson}, \bibinfo{person}{P. Ammirato},
  \bibinfo{person}{C.~Y. Fu}, \bibinfo{person}{W. Liu}, \bibinfo{person}{J.
  Kosecka}, {and} \bibinfo{person}{A.~C. Berg}.}
  \bibinfo{year}{2016}\natexlab{}.
\newblock \showarticletitle{Fast Single Shot Detection and Pose Estimation}. In
  \bibinfo{booktitle}{{\em 3DV}}. \bibinfo{pages}{676--684}.
\newblock


\bibitem[\protect\citeauthoryear{Ren, He, Girshick, and Sun}{Ren
  et~al\mbox{.}}{2017}]%
        {Ren:2017:FRT}
\bibfield{author}{\bibinfo{person}{Shaoqing Ren}, \bibinfo{person}{Kaiming He},
  \bibinfo{person}{Ross Girshick}, {and} \bibinfo{person}{Jian Sun}.}
  \bibinfo{year}{2017}\natexlab{}.
\newblock \showarticletitle{Faster R-CNN: Towards Real-Time Object Detection
  with Region Proposal Networks}.
\newblock \bibinfo{journal}{{\em {IEEE PAMI}\/}} \bibinfo{volume}{39},
  \bibinfo{number}{6} (\bibinfo{year}{2017}), \bibinfo{pages}{1137--1149}.
\newblock


\bibitem[\protect\citeauthoryear{Rogez, Weinzaepfel, and Schmid}{Rogez
  et~al\mbox{.}}{2018}]%
        {RogezWS18}
\bibfield{author}{\bibinfo{person}{Gr\'egory Rogez}, \bibinfo{person}{Philippe
  Weinzaepfel}, {and} \bibinfo{person}{Cordelia Schmid}.}
  \bibinfo{year}{2018}\natexlab{}.
\newblock \showarticletitle{{LCR-Net++: Multi-person 2D and 3D Pose Detection
  in Natural Images}}.
\newblock \bibinfo{journal}{{\em {CoRR}\/}}  \bibinfo{volume}{abs/1803.00455v1}
  (\bibinfo{year}{2018}).
\newblock


\bibitem[\protect\citeauthoryear{Satkin and Hebert}{Satkin and Hebert}{2013}]%
        {Satkin2013}
\bibfield{author}{\bibinfo{person}{S. Satkin} {and} \bibinfo{person}{M.
  Hebert}.} \bibinfo{year}{2013}\natexlab{}.
\newblock \showarticletitle{3DNN: Viewpoint Invariant 3D Geometry Matching for
  Scene Understanding}. In \bibinfo{booktitle}{{\em {IEEE CVPR}}}.
  \bibinfo{pages}{1873--1880}.
\newblock
\showISSN{1550-5499}
\showDOI{%
\url{https://doi.org/10.1109/ICCV.2013.235}}


\bibitem[\protect\citeauthoryear{Savva, Chang, Hanrahan, Fisher, and
  Nie{\ss}ner}{Savva et~al\mbox{.}}{2014}]%
        {savva2014scenegrok}
\bibfield{author}{\bibinfo{person}{Manolis Savva}, \bibinfo{person}{Angel~X.
  Chang}, \bibinfo{person}{Pat Hanrahan}, \bibinfo{person}{Matthew Fisher},
  {and} \bibinfo{person}{Matthias Nie{\ss}ner}.}
  \bibinfo{year}{2014}\natexlab{}.
\newblock \showarticletitle{SceneGrok: Inferring Action Maps in 3D
  Environments}.
\newblock \bibinfo{journal}{{\em {ACM SIGGRAPH Asia}\/}} \bibinfo{volume}{33},
  \bibinfo{number}{6} (\bibinfo{year}{2014}).
\newblock


\bibitem[\protect\citeauthoryear{Savva, Chang, Hanrahan, Fisher, and
  Nie{\ss}ner}{Savva et~al\mbox{.}}{2016}]%
        {savva2016pigraphs}
\bibfield{author}{\bibinfo{person}{Manolis Savva}, \bibinfo{person}{Angel~X.
  Chang}, \bibinfo{person}{Pat Hanrahan}, \bibinfo{person}{Matthew Fisher},
  {and} \bibinfo{person}{Matthias Nie{\ss}ner}.}
  \bibinfo{year}{2016}\natexlab{}.
\newblock \showarticletitle{{PiGraphs: Learning Interaction Snapshots from
  Observations}}.
\newblock \bibinfo{journal}{{\em {ACM SIGGRAPH}\/}} \bibinfo{volume}{35},
  \bibinfo{number}{4} (\bibinfo{year}{2016}).
\newblock


\bibitem[\protect\citeauthoryear{Schwing, Fidler, Pollefeys, and
  Urtasun}{Schwing et~al\mbox{.}}{2013}]%
        {Schwing2013}
\bibfield{author}{\bibinfo{person}{A.~G. Schwing}, \bibinfo{person}{S. Fidler},
  \bibinfo{person}{M. Pollefeys}, {and} \bibinfo{person}{R. Urtasun}.}
  \bibinfo{year}{2013}\natexlab{}.
\newblock \showarticletitle{Box in the Box: Joint 3D Layout and Object
  Reasoning from Single Images}. In \bibinfo{booktitle}{{\em {IEEE ICCV}}}.
  \bibinfo{pages}{353--360}.
\newblock
\showISSN{1550-5499}
\showDOI{%
\url{https://doi.org/10.1109/ICCV.2013.51}}


\bibitem[\protect\citeauthoryear{Shao, Xu, Zhou, Wang, Li, and Guo}{Shao
  et~al\mbox{.}}{2012}]%
        {Shao:2012}
\bibfield{author}{\bibinfo{person}{Tianjia Shao}, \bibinfo{person}{Weiwei Xu},
  \bibinfo{person}{Kun Zhou}, \bibinfo{person}{Jingdong Wang},
  \bibinfo{person}{Dongping Li}, {and} \bibinfo{person}{Baining Guo}.}
  \bibinfo{year}{2012}\natexlab{}.
\newblock \showarticletitle{An Interactive Approach to Semantic Modeling of
  Indoor Scenes with an RGBD Camera}.
\newblock \bibinfo{journal}{{\em {ACM SIGGRAPH Asia}\/}} \bibinfo{volume}{31},
  \bibinfo{number}{6}, Article \bibinfo{articleno}{136} (\bibinfo{date}{Nov.}
  \bibinfo{year}{2012}), \bibinfo{numpages}{11}~pages.
\newblock
\showISSN{0730-0301}


\bibitem[\protect\citeauthoryear{Tekin, Rozantsev, Lepetit, and Fua}{Tekin
  et~al\mbox{.}}{2016}]%
        {Tekin2016}
\bibfield{author}{\bibinfo{person}{Bugra Tekin}, \bibinfo{person}{Artem
  Rozantsev}, \bibinfo{person}{Vincent Lepetit}, {and} \bibinfo{person}{Pascal
  Fua}.} \bibinfo{year}{2016}\natexlab{}.
\newblock \showarticletitle{Direct Prediction of 3D Body Poses from Motion
  Compensated Sequences}. In \bibinfo{booktitle}{{\em {IEEE CVPR}}}.
\newblock


\bibitem[\protect\citeauthoryear{Tome, Russell, and Agapito}{Tome
  et~al\mbox{.}}{2017}]%
        {tome2017lifting}
\bibfield{author}{\bibinfo{person}{Denis Tome}, \bibinfo{person}{Chris
  Russell}, {and} \bibinfo{person}{Lourdes Agapito}.}
  \bibinfo{year}{2017}\natexlab{}.
\newblock \showarticletitle{Lifting from the Deep: Convolutional 3D Pose
  Estimation from a Single Image}.
\newblock \bibinfo{journal}{{\em {IEEE CVPR}\/}} (\bibinfo{year}{2017}).
\newblock


\bibitem[\protect\citeauthoryear{Toshev and Szegedy}{Toshev and
  Szegedy}{2014}]%
        {Toshev:2014}
\bibfield{author}{\bibinfo{person}{Alexander Toshev} {and}
  \bibinfo{person}{Christian Szegedy}.} \bibinfo{year}{2014}\natexlab{}.
\newblock \showarticletitle{DeepPose: Human Pose Estimation via Deep Neural
  Networks}. In \bibinfo{booktitle}{{\em {IEEE CVPR}}}.
\newblock
\showISBNx{978-1-4799-5118-5}


\bibitem[\protect\citeauthoryear{von Marcard, Rosenhahn, Black, and
  Pons-Moll}{von Marcard et~al\mbox{.}}{2017}]%
        {vonMarcard:2017}
\bibfield{author}{\bibinfo{person}{T. von Marcard}, \bibinfo{person}{B.
  Rosenhahn}, \bibinfo{person}{M.~J. Black}, {and} \bibinfo{person}{G.
  Pons-Moll}.} \bibinfo{year}{2017}\natexlab{}.
\newblock \showarticletitle{Sparse Inertial Poser: Automatic 3D Human Pose
  Estimation from Sparse IMUs}.
\newblock \bibinfo{journal}{{\em {CGF Eurographics}\/}} \bibinfo{volume}{36},
  \bibinfo{number}{2} (\bibinfo{date}{May} \bibinfo{year}{2017}),
  \bibinfo{pages}{349--360}.
\newblock
\showISSN{0167-7055}


\bibitem[\protect\citeauthoryear{Wei, Zhao, Zheng, and Zhu}{Wei
  et~al\mbox{.}}{2013}]%
        {Wei:2013}
\bibfield{author}{\bibinfo{person}{Ping Wei}, \bibinfo{person}{Yibiao Zhao},
  \bibinfo{person}{Nanning Zheng}, {and} \bibinfo{person}{Song-Chun Zhu}.}
  \bibinfo{year}{2013}\natexlab{}.
\newblock \showarticletitle{Modeling 4D Human-Object Interactions for Event and
  Object Recognition}. In \bibinfo{booktitle}{{\em {IEEE ICCV}}}.
\newblock
\showISBNx{978-1-4799-2840-8}


\bibitem[\protect\citeauthoryear{Wei, Ramakrishna, Kanade, and Sheikh}{Wei
  et~al\mbox{.}}{2016}]%
        {wei2016}
\bibfield{author}{\bibinfo{person}{Shih-En Wei}, \bibinfo{person}{Varun
  Ramakrishna}, \bibinfo{person}{Takeo Kanade}, {and} \bibinfo{person}{Yaser
  Sheikh}.} \bibinfo{year}{2016}\natexlab{}.
\newblock \showarticletitle{Convolutional pose machines}. In
  \bibinfo{booktitle}{{\em {IEEE CVPR}}}.
\newblock


\bibitem[\protect\citeauthoryear{Xu, Ma, Zhang, Zhu, Shamir, Cohen-Or, and
  Huang}{Xu et~al\mbox{.}}{2014}]%
        {Xu:2014}
\bibfield{author}{\bibinfo{person}{Kai Xu}, \bibinfo{person}{Rui Ma},
  \bibinfo{person}{Hao Zhang}, \bibinfo{person}{Chenyang Zhu},
  \bibinfo{person}{Ariel Shamir}, \bibinfo{person}{Daniel Cohen-Or}, {and}
  \bibinfo{person}{Hui Huang}.} \bibinfo{year}{2014}\natexlab{}.
\newblock \showarticletitle{Organizing Heterogeneous Scene Collections Through
  Contextual Focal Points}.
\newblock \bibinfo{journal}{{\em {ACM SIGGRAPH}\/}} \bibinfo{volume}{33},
  \bibinfo{number}{4}, Article \bibinfo{articleno}{35} (\bibinfo{date}{July}
  \bibinfo{year}{2014}), \bibinfo{numpages}{12}~pages.
\newblock
\showISSN{0730-0301}


\bibitem[\protect\citeauthoryear{Yao, Khosla, and Fei-Fei}{Yao
  et~al\mbox{.}}{2011}]%
        {icml11_yao}
\bibfield{author}{\bibinfo{person}{Bangpeng Yao}, \bibinfo{person}{Aditya
  Khosla}, {and} \bibinfo{person}{Li Fei-Fei}.}
  \bibinfo{year}{2011}\natexlab{}.
\newblock \showarticletitle{Classifying Actions and Measuring Action Similarity
  by Modeling the Mutual Context of Objects and Human Poses}. In
  \bibinfo{booktitle}{{\em ICML}}.
\newblock


\bibitem[\protect\citeauthoryear{Yeh, Yang, Watson, Goodman, and Hanrahan}{Yeh
  et~al\mbox{.}}{2012}]%
        {Yeh:2012}
\bibfield{author}{\bibinfo{person}{Yi-Ting Yeh}, \bibinfo{person}{Lingfeng
  Yang}, \bibinfo{person}{Matthew Watson}, \bibinfo{person}{Noah~D. Goodman},
  {and} \bibinfo{person}{Pat Hanrahan}.} \bibinfo{year}{2012}\natexlab{}.
\newblock \showarticletitle{Synthesizing Open Worlds with Constraints Using
  Locally Annealed Reversible Jump MCMC}.
\newblock \bibinfo{journal}{{\em {ACM SIGGRAPH}\/}} \bibinfo{volume}{31},
  \bibinfo{number}{4}, Article \bibinfo{articleno}{56} (\bibinfo{date}{July}
  \bibinfo{year}{2012}), \bibinfo{numpages}{11}~pages.
\newblock
\showISSN{0730-0301}


\bibitem[\protect\citeauthoryear{Zhang, Lei, Zhong, Du, and Peng}{Zhang
  et~al\mbox{.}}{2016}]%
        {zhang16}
\bibfield{author}{\bibinfo{person}{Hong-Bo Zhang}, \bibinfo{person}{Qing Lei},
  \bibinfo{person}{Bi-Neng Zhong}, \bibinfo{person}{Ji-Xiang Du}, {and}
  \bibinfo{person}{JiaLin Peng}.} \bibinfo{year}{2016}\natexlab{}.
\newblock \showarticletitle{A Survey on Human Pose Estimation}.
\newblock \bibinfo{journal}{{\em Intelligent Automation and Soft Computing\/}}
  \bibinfo{volume}{22}, \bibinfo{number}{3} (\bibinfo{year}{2016}),
  \bibinfo{pages}{483--489}.
\newblock


\bibitem[\protect\citeauthoryear{Zhao, Hu, Guerrero, Mitra, and Komura}{Zhao
  et~al\mbox{.}}{2016}]%
        {Zhao:2016}
\bibfield{author}{\bibinfo{person}{Xi Zhao}, \bibinfo{person}{Ruizhen Hu},
  \bibinfo{person}{Paul Guerrero}, \bibinfo{person}{Niloy Mitra}, {and}
  \bibinfo{person}{Taku Komura}.} \bibinfo{year}{2016}\natexlab{}.
\newblock \showarticletitle{Relationship Templates for Creating Scene
  Variations}.
\newblock \bibinfo{journal}{{\em {ACM SIGGRAPH Asia}\/}} \bibinfo{volume}{35},
  \bibinfo{number}{6}, Article \bibinfo{articleno}{207} (\bibinfo{date}{Nov.}
  \bibinfo{year}{2016}), \bibinfo{numpages}{13}~pages.
\newblock
\showISSN{0730-0301}


\bibitem[\protect\citeauthoryear{Zhou, Zhu, Leonardos, Derpanis, and
  Daniilidis}{Zhou et~al\mbox{.}}{2016}]%
        {Zhou:Monocap:2015}
\bibfield{author}{\bibinfo{person}{Xiaowei Zhou}, \bibinfo{person}{Menglong
  Zhu}, \bibinfo{person}{Spyridon Leonardos}, \bibinfo{person}{Kosta Derpanis},
  {and} \bibinfo{person}{Kostas Daniilidis}.} \bibinfo{year}{2016}\natexlab{}.
\newblock \showarticletitle{Sparseness Meets Deepness: 3D Human Pose Estimation
  from Monocular Video}.
\newblock \bibinfo{journal}{{\em {IEEE CVPR}\/}} (\bibinfo{year}{2016}).
\newblock


\end{thebibliography}
